\documentclass[usenatbib,usegraphicx,referee]{mn2e}
\def\xb{\bar{x}_{\rm HI}}
\def\B{B_{\nu}}
\def\k{{\bf k}}
\def\HI{{\rm HI}}

\def\physrep{PHYS REP}
\def\pasp{PASP}
\def\mnras{MNRAS}
\def\apj{Ap.J}
\def\apjs{Ap.JS}

\def\u{{\bf U}} 

\def\V{\mathcal{V}}

\def\del{\partial}
\usepackage{graphics}
\begin{document}
\date {}
\title[GMRT   observation towards detecting 21-cm
  signal]   
{GMRT   observation towards detecting the Post-reionization 21-cm
  signal}   

\author[A. Ghosh, S. Bharadwaj, S. S. Ali and J. N. Chengalur]{Abhik
  Ghosh$^{1}$\thanks{E-mail: abhik@phy.iitkgp.ernet.in}, Somnath
  Bharadwaj$^{1}$\thanks{Email:somnath@phy.iitkgp.ernet.in},
  Sk. Saiyad Ali$^{2}$\thanks{Email:saiyad@phys.jdvu.ac.in} and \and
  Jayaram N. Chengalur$^{3}$\thanks{Email:chengalu@ncra.tifr.res.in}
  \\ $^{1}$ Department of Physics and Meteorology \& Centre for
  Theoretical Studies , IIT Kharagpur, 721 302 , India \\$^{2}$
  Department of Physics,Jadavpur University, Kolkata 700032, India.
  \\$^3$ National Centre for Radio Astrophysics, TIFR, Post Bag 3,
  Ganeshkhind, Pune 411 007, India}

\maketitle
   
\begin{abstract}
The redshifted $21$-cm signal  from  neutral hydrogen (HI) is an
important future probe of  the the high redshift universe. We have analyzed
$610 \ {\rm MHz}$ GMRT observations towards detecting this signal from 
$z=1.32$. The   multi-frequency angular power spectrum
 $C_{\ell}(\Delta \nu)$ is used to  characterize
 the statistical properties of the background radiation across angular
 scales $\sim 20^{''}$ to $10^{'}$,  and a  frequency bandwidth of
 $7.5\; \rm MHz$ with  resolution  $125\;  \rm  kHz$. The measured
 $C_{\ell}(\Delta \nu)$ which ranges from  $7 \ {\rm mK}^2$ to 
$18  \ {\rm mK^2}$ is dominated by foregrounds, the expected HI signal
 $C^{\rm HI}_{\ell}(\Delta \nu) \sim 10^{-6} - 10^{-7} \ {\rm mK}^2$
 is several orders of magnitude smaller and detecting this is a big
 challenge. The foregrounds,
 believed to originate from continuum sources, is expected to vary
 smoothly with $\Delta \nu$ whereas the HI signal decorrelates within
 $\sim 0.5 \ {\rm MHz}$ and this holds the promise of separating the
 two. For each $\ell$, we  use the interval $0.5 \le \Delta
 \nu \le 7.5 \ {\rm MHz}$ to fit a fourth order polynomial
 which is subtracted from   the measured $C_{\ell}(\Delta \nu)$ 
to remove any smoothly varying component across the entire bandwidth 
$\Delta \nu \le 7.5 \ {\rm MHz}$.  The residual $C_{\ell}(\Delta
\nu)$, we find, has an oscillatory pattern with amplitude and period
respectively $\sim 0.1 \ {\rm mK}^2$ and  $\Delta \nu = 3 \ {\rm MHz}$
at the smallest $\ell$ value of $1476$, and  the amplitude and period
decreasing with increasing $\ell$. Applying a suitably chosen high
pass filter, we are able to remove the residual oscillatory pattern
for $\ell=1476$ where the residual $C_{\ell}(\Delta \nu)$ is now
consistent with zero at the $3\sigma$ noise level. Based on this we
conclude that we have successfully removed the foregrounds at
$\ell=1476$ and the residuals are consistent with noise. We use 
this to place an upper limit on the HI signal whose amplitude is
determined by $\xb b$ $( C^{\rm HI}_{\ell}(\Delta \nu) \propto [\xb
  b]^2)$, where $\xb$ and $b$ are the HI neutral fraction and the HI
bias respectively.  A value of $\xb b$ greater than $7.95$ would have
been detected in our observation, and is therefore  ruled out 
at the $3\sigma$ level. 
For comparison, studies of quasar absorption spectra indicate $\xb
\sim 2.5 \times 10^{-2}$ which is $\sim 330$ times smaller than our
upper limit. We have not succeeded in completely removing the residual
oscillatory pattern, whose cause is presently unknown to us, for the
larger $\ell$  values. 
\end{abstract} 


\section{Introduction}
Detecting redshifted 21-cm radiation from neutral hydrogen (HI) at
high redshifts is of considerable interest in cosmology.  At redshifts
$z \le 6$, the bulk of the neutral gas is in clouds that have HI
column densities in excess of $2 \times 10^{20}\,\,{\rm atoms/cm^{2}}$
\citep{lanzetta,lombardi,peroux}. These high column density clouds are
observed as damped Lyman-$\alpha$ absorption lines seen in quasar
spectra. The analysis of quasar spectra indicate that the ratio of the
density $\rho_{\rm gas}(z)$ of neutral gas to the present critical
density $\rho_{\rm co}$ of the universe has a nearly constant value $
\Omega_{\rm gas}(z) \sim \rho_{\rm gas}(z)/\rho_{\rm co} \sim
10^{-3}$, over a large redshift range $0.5 \le z \le 5.0$
\citep*{lombardi,Rao,peroux,prochaska,rao,kanekar}.  The redshifted
$21 \, {\rm cm}$ radiation from the HI in this redshift range will be
seen in emission.  The emission from individual clouds ($ < 10
\,\mu{\rm Jy}$) is too weak to be detected with existing instruments
unless the image is significantly magnified by gravitational lensing
\citep{saini}. The collective emission from the undetected clouds is
present as a very faint background in all radio observations at
frequencies below $1420 \, {\rm MHz}$. The fluctuations in this
background radiation carry an imprint of the HI distribution at the
redshift $z$ where the radiation originated.  The possibility of
detecting this holds the potential of providing us with a new
observational probe of large-scale structures
\citep{kum,bagla1,BNS,BSS,BP3,BS4,Ali,wyithe1,Wyithe,Bagla,NKB}.  In a
recent paper \citet{pen} report a detection of the post-reionization HI
signal through the cross-correlation between the HIPASS and the 6dfGRS
data. 

Observations of redshifted 21-cm radiation can in principle be carried
out over a large redshift range starting from the cosmological Dark
Ages through the Epoch of Reionization to the present epoch
\citep{BA5}, allowing us to trace out both the evolution history of
neutral hydrogen as well as the growth of structures in the
universe.  Redshifted 21-cm observations also hold the potential of
allowing us to probe the expansion history of the universe
\citep*{mcquinn,chang,visbal,BST}.

 The Giant Meter Wave Radio Telescope (GMRT
 \footnote{http://www.gmrt.ncra.tifr.res.in}; \citealt{swarup}),
 currently operating at several frequency bands in the frequency range
 $150$ to  $ 1420 \, {\rm MHz}$ is well suited for carrying out 
 observations towards detecting the HI signal over a large  
redshift range from $z \sim 0$  to  $z \sim 8.3$ and angular scales of
 $\sim 10^{''}$ to $\sim 1^{\circ}$. 
 In this paper we
 report results from the analysis of $610 \, {\rm MHz}$ observations
 towards detecting the redshifted 21-cm signal from the cosmological
 HI distribution at $z = 1.32$.

We have characterized, possibly for the first
time,  the statistical properties of the background radiation at $610
\, {\rm MHz}$ across $\sim 20^{''}$ to $10^{'}$ angular scales and a
frequency bandwidth of $7.5\; \rm MHz$ with a resolution of $125\; \rm
kHz$ using  the  multi-frequency angular power spectrum
$C_{\ell}(\Delta \nu)$ (hereafter MAPS;  \citealt{kanan}).
This  jointly characterizes the  angular ($\ell$)
and frequency ($\Delta \nu$) dependence of the fluctuations in the 
$610 \, {\rm MHz}$ radiation in the field of view of our observation.  
Foregrounds from different astrophysical sources are expected to be a
few orders of magnitude larger than  the predicted 21-cm signal 
\citep*{Shaver,dmat,Oh,Santos,wang,ali} and our  $610 \, {\rm MHz}$
GMRT observations  
are expected  to be nearly entirely  dominated by foregrounds
which are predicted to be at least a thousand times larger than the HI
signal.
Separating the HI signal from foregrounds is the most
important challenge for cosmological  redshifted 21-cm observations.

The foregrounds are believed  to have a smooth continuum spectra and 
the contribution to $C_{\ell}(\Delta \nu)$ is expected to vary very
slowly with $\Delta \nu$ across the band $(7.5 \, {\rm MHz})$ of our
analysis. The contribution from the HI signal decorrelates very
rapidly with increasing $\Delta \nu$ and is expected to be
uncorrelated beyond  $\Delta \nu = 0.5 \, {\rm MHz}$ at the angular
scales ($\ell=10^3$ to $\ell=3 \times 10^4$) of our analysis. 
 This property of the signal holds the promise of allowing
us to separate the signal from the foregrounds. In this paper we 
propose and implement a technique that uses polynomial fitting 
in $\Delta \nu$ to subtract out any smoothly varying component from
the measured  $C_{\ell}(\Delta \nu)$. The residuals are
expected to contain only the HI signal and noise. The target of the
present work  is to test if the polynomial subtraction successfully removes
the foregrounds to a level such that the residuals are consistent with
noise.   The noise in the current 
observation is considerably larger than the HI-signal and longer
observations would be needed for detecting the HI signal. 

The present work closely follows an earlier paper \citep{ali} which
analyzed $150 \, {\rm MHz}$ GMRT observations.  We note that the
prospect of detecting the redshifted 21-cm signal considerably
increases at higher frequencies (eg. $610 \, {\rm MHz}$) where the
foreground contribution and noise are both smaller. Further, the
problem of man made radio frequency interference is considerably more
severe at $150 \, {\rm MHz}$ as compared to $610 \, {\rm MHz}$.

A  brief outline of the paper follows. Section $2$ 
describes the observation and data analysis,  Section $3$
presents the  visibility correlation technique that we use to estimate 
$C_{\ell}(\Delta \nu)$  and also presents the estimated values,  
Sections $4$ and $5$ present model predictions for 
the HI signal and foregrounds respectively, while Section $6$ 
describes our proposed technique of foreground removal and finally
Section $7$  contains results and conclusions. 

\section{GMRT Observations and Data Analysis}

The GMRT has $30$ fixed antennas each of diameter $45 \, \rm m$. $14$
of which  are randomly distributed in a central square 
$1.1 \, \rm km\times1.1 \, \rm km$ in extent, while the 
rest  of the antennas are distributed approximately  in a 'Y' shaped
configuration. The shortest antenna separation (baseline) is around
$60 \, \rm m$ including  projection effects while the
largest separation  can be as long as $26 \, \rm km$. The hybrid
configuration of the GMRT gives reasonably 
good sensitivity to probe both compact and extended sources.

The observed field of view is centered on
$\alpha_{2000}=12^h36^m49^s$, $\delta_{2000}=
62^{\circ}17^{'}57^{''}\, $ which is situated near Hubble Deep Field
North (HDF-N) ($\alpha_{2000}=12^h36^m49.4^s,\delta_{2000}=
62^{\circ}12^{'}58^{''}\, $). The galactic co-ordinates of the
observed field is $l=125.87^{o}, b = 54.74^{o}$. The sky temperature
determined at this location is 20 K in the 408 MHz \cite {haslam}
map. The observations were carried out over three days from $4^{\rm
  th}$ to $7^{\rm th}$ September, $2002$ and the total observation
time was almost 30 hours (including calibration). The observation had
a center frequency of $618 \, {\rm MHz}$, and a total band-width of
$16 \, {\rm MHz}$ divided into $128$ frequency channels, each $125 \,{\rm
  kHz}$ wide. The integration time was 16 seconds and visibilities
were recorded for two orthogonal circular polarizations. The
calibrator source 3C147 and 3C286 were used for flux calibration and
1313+675 was used for phase calibration.  The phase calibrator was
observed every half hour to correct for temporal variations in the
system gain.  We have used the Astronomical Image Processing Software
(AIPS) to analyze the recorded visibility data.  The flux of these two
flux calibrator was estimated by extending the Baars scale
(\citealt{Baars}) to low frequencies using the AIPS task
'SETJY'. Standard AIPS tasks were used to flag all data that could be
visually identified as being bad. The entire lower sideband data was
found to be bad and was discarded from the subsequent analysis.  Data
from different days were calibrated and flagged separately and then
combined using the AIPS task 'DBCON'. We find that the channels near
the edge of the band are relatively noisy and hence only the 100
central channels were used in the subsequent analysis.

An initial 2D image of the field of view (FOV) showed four bright
sources with considerable imaging artifacts.  To improve our image
quality , initially we have subtracted out the clean components (CC)
of these bright sources by moving them to the phase center using
appropriate RA-SHIFT and DEC-SHIFT within AIPS. Then, we add back the
brightest source and use this for three rounds of self-calibration
with time intervals of 3 and 2 minutes for phase calibration and
finally 20 minutes for amplitude and phase calibration and then
subtract out the brightest source again. The same process is followed
for rest of the bright sources. Subsequent to this, we have also
subtracted out all the weaker sources from our FOV and used the AIPS
task 'TVFLG' to flag out any bad visibility. Finally, we have
collapsed all frequency channels and clipped the resulting
visibilities at $0.07\, {\rm Jy}$.  At each stage the same calibration
and flag tables were also applied to the original 100 channel data
which contains all the sources.
 
\begin{figure}
\begin{centering}
\includegraphics[width=90mm]{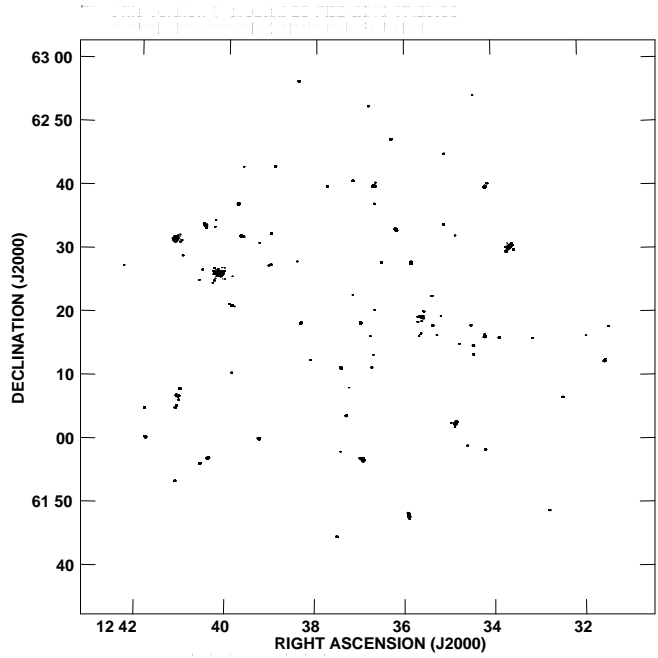}
\caption{Our continuum image of bandwidth $12.5 \, {\rm
    MHz}$ centered at $618 \, {\rm MHz}$. The FOV was imaged
  using $163$ facets which have been combined using the AIPS
  task FLATN. The rms off-source noise is $ 60 \,{\rm \mu
    Jy/Beam}$. Most  of the extended features visible in the image 
  are imaging artifacts around the four bright sources.}
\label{fig:con1}
\end{centering} 
\end{figure}

 The large field of view ($\theta_{\rm FWHM}=43'$) of the GMRT at 610
 MHz leads to considerable errors if the non-planar nature of the GMRT
 antenna distribution is not taken into account. We use the three
 dimensional (3D) imaging feature \citep{perley} in the AIPS task
 'IMAGR' in which the entire field of view is divided into multiple
 sub-fields, each of which is imaged separately. Here, a
 $1.5^{\circ}\times1.5^{\circ}$ FOV was imaged using 163 facets. The
 presence of a large number of sources in the field allows us to do
 self calibration loops to improve the image quality. We have applied
 4 rounds of self calibration, the first three only for the phase and
 the final round for both amplitude and phase.  The time interval for
 the gain correction was chosen as $3, 2, 1 $ and $20$ \,minutes for
 the successive self calibration loops. At every stage the calibration
 tables were applied to the original 100 channel data. The 100
 channels were collapsed into 10 channels which were used to make a
 continuum image of the entire FOV. Figure \ref{fig:con1} shows our
 continuum image of band width $12.5 \, {\rm MHz}$ centered at $617 \,
 {\rm MHz}$. The synthesized beam has FWHM $8.2^{''}\times 5.8^{''} $.
 The off source rms. noise in the image was $\sim 60 \, {\mu}{\rm
   Jy/Beam}$ and the image quality had improved considerably. The rms
 noise around the brighter sources is higher, using the AIPS task
 'TVSTAT' we notice that it is around $100\, {\mu}{\rm Jy/Beam}$. The
 brighter sources are also found to be accompanied by a region of
 negative flux density, these are presumably the results of residual
 phase errors which were not corrected in our self calibration
 process.  The maximum and minimum flux density in the final image are
 $ \sim 250 \,{\rm mJy/Beam}$ and $\sim -2.8 \, {\rm mJy/Beam}$
 respectively.

The subsequent analysis was done using the calibrated visibilities of
the original 100 channel data with all the sources. The data contains
$510528$ baselines, each of which has visibilities for $2$ circular
polarizations. The baselines are in the range $ 200 \, \rm \lambda$ to
$ 20 \, \rm k\lambda$. The visibilities from the two polarizations
were combined ($\V=[\V_{RR}+\V_{LL}]/2$) for the subsequent
analysis. The final calibrated data contains $33233698$
visibilities. The real part of the visibilities have a mean of $0.68
\, {\rm mJy}$ and rms of $0.25\,{\rm Jy}$. Similarly, the imaginary
part has a mean of $2.34 \, {\rm mJy}$ and rms of $0.25\,{\rm Jy}$.

\begin{figure}
\begin{centering}
\includegraphics[width=55mm,angle=-90]{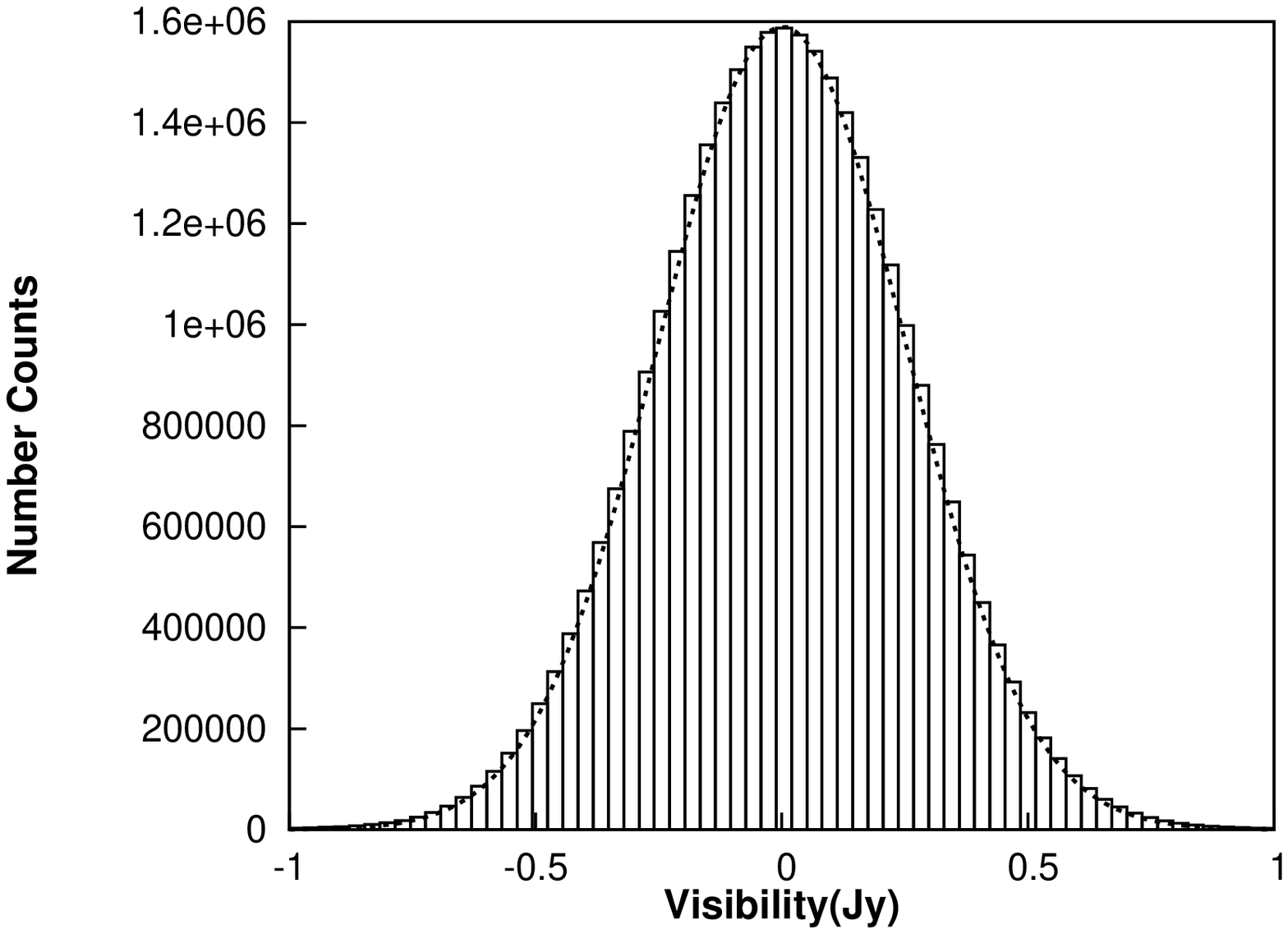}
\includegraphics[width=55mm,angle=-90]{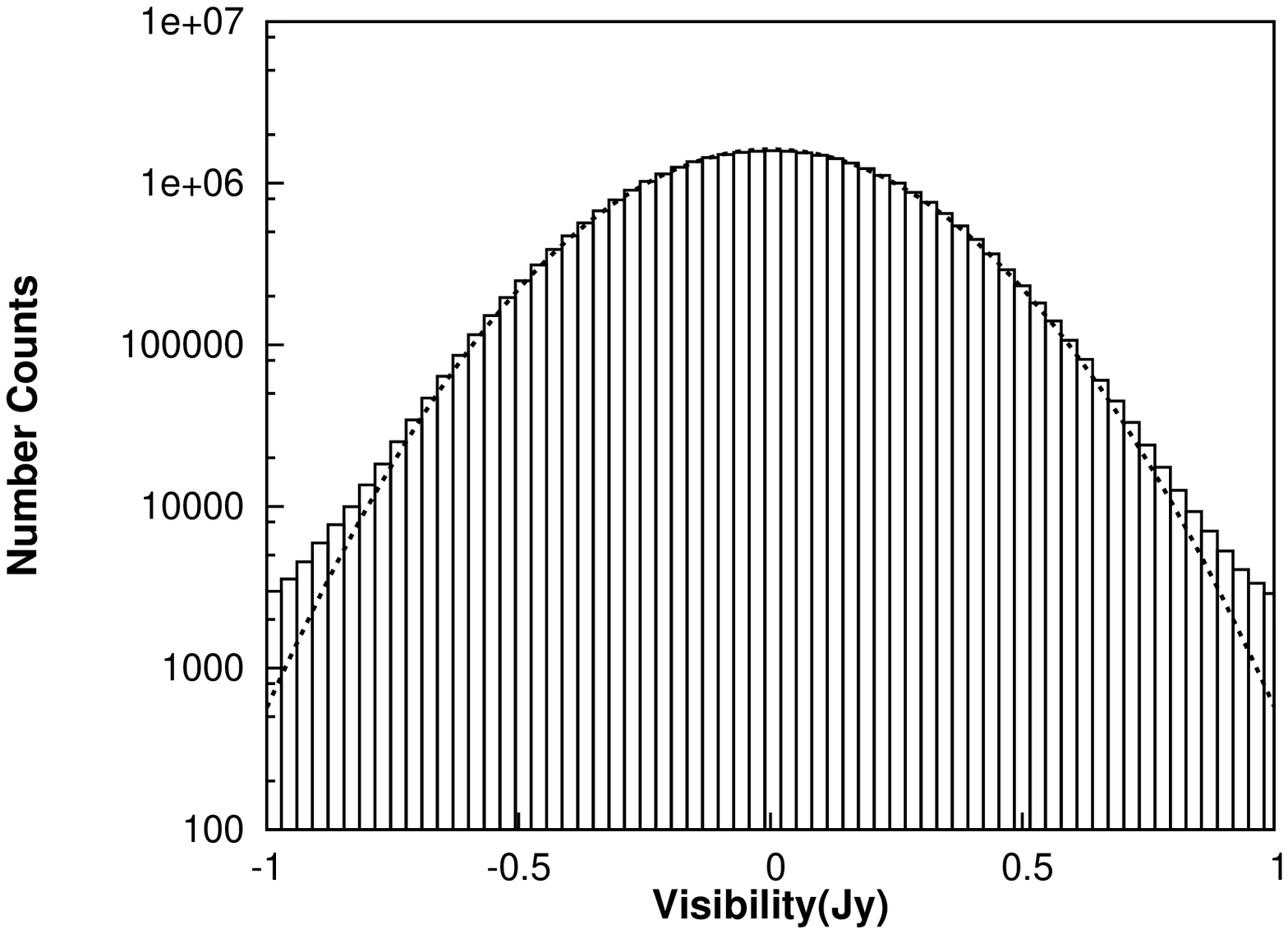}
\caption{Histogram showing  the distribution of the real parts of all
the $33233698$ measured visibilities. The same is shown on a linear scale
(left panel) and a log-linear scale (right panel). A   Gaussian with
mean $0.683 \, {\rm  mJy}$ and rms $0.251\,{\rm Jy}$, values 
calculated for the real parts  of the measured visibilities,  is
plotted as a dotted line in both panels. While the Gaussian fits the
data very well at small amplitudes, there is a 
 discrepancy at high
amplitudes ($\sim 0.75 {\rm Jy}$) which is visible only in the right
panel.} 
\label{fig:h1}
\end{centering}
\end{figure}

It is often convenient to assume that the visibilities have a Gaussian
distribution. The distribution of the real part of the visibilities is
shown in a histogram (Figure \ref{fig:h1}). We find that a
Gaussian gives a reasonably good fit to the data within $3 \sigma$,
which contains the bulk of the data. The number counts predicted by
the Gaussian falls much faster than the data at large visibility
values $\mid Re(\V) \mid > 0.75 \, {\rm Jy}$. The imaginary part of the
visibilities show a similar behavior.

 Deviation from 
Gaussian statistics is  expected to mainly effect the error estimate in the visibility correlation. We expect this effect to be small , since the
discrepancy is for only small fraction of visibilities.

\section{ Visibility correlations and the angular power spectrum}
\label{sec3}
The aim here is to quantify the statistical properties, in angle and
frequency,  of the $610 \ {\rm MHz}$ sky signal. For a frequency $\nu$, 
the angular dependence of the   brightness temperature   distribution
on the sky $T(\nu,{\bf   \hat{n}})$
 may be  expanded   in spherical harmonics as  
\begin{equation}
T(\nu,{\bf \hat{n}}) = \sum_{\ell,m} \
a_{\ell m}(\nu) \ Y_{\ell m} ({\bf \hat{n}})\,.
\label{eq:t1}
\end{equation}
The   multi-frequency angular power spectrum (MAPS; 
\citealt{kanan}),  which  jointly characterizes the  dependence on
angular scale  and frequency separation, is defined as 
\begin{equation}
C_l(\Delta \nu) \equiv C_l(\nu, \nu + \Delta \nu)=\langle a_{lm}(\nu)
~ a^*_{lm}(\nu + \Delta \nu) \rangle \,.
\label{eq:c1}
\end{equation} 
Here $\ell$ refers to the angular modes on the sky. The sky signal is
assumed to 
be statistically isotropic.  We also assume that for the relatively small 
bandwidth of our observation  ($\Delta \nu \ll \nu$),   
the frequency dependence can be entirely characterized   through
$\Delta \nu$ whereby  we do not explicitly show $\nu$ as an argument
in eq. (\ref{eq:c1}).

We use the  correlation between pairs of  visibilities   $\V(\u,\nu)$
and  $\V(\u+\Delta \u,\nu+ \Delta \nu)$ 
\begin{equation}
V_2(\u,\nu;\u + \Delta \u,\nu+\Delta \nu) \equiv \langle \V(\u,\nu) 
\V^*(\u+\Delta \u,\nu+ \Delta \nu)  \rangle
\label{eq:v2}
\end{equation}
to estimate $C_{\ell}(\Delta \nu)$. 
 The correlation of a visibility
with itself is excluded to avoid a positive noise bias in the
estimator. 
 \citet{ali} as well as \citet{Dutta} contain detailed
discussions of the estimator and we highlight only a few salient
features here. 

The GMRT primary beam pattern is well approximated by a Gaussian
$A(\theta)=e^{-\theta^2/\theta_0^2}$ where $\theta_0 \approx 0.6
  \times \theta_{\rm FWHM}=25^{'}.8$  ($\theta_{\rm FWHM}=43^{'}$)
at 610 MHz. In a situation
  where $\Delta \u$ is small such  that     
$\mid \Delta \u \mid < (\pi \theta_0)^{-1}=42.4 \ \lambda$
  ($\theta_0$ in radians),   the expected correlation 
$V_2(\u,\nu;\u + \Delta \u,\nu+\Delta \nu)$ in 
  eq. (\ref{eq:v2})  does not depend on $\Delta \u$ whereby we may 
express it as $V_2(U,\Delta \nu)$.  Further, if $U \gg
  (\pi \theta_0)^{-1}$ we have

\begin{equation}
V_2(U,\Delta \nu)=\frac{\pi\theta_0^2}{2}
 \left(\frac{\del \B}{\del   T}\right)^2
C_{2\pi U}(\Delta \nu) \,Q(\Delta \nu) \,.
\label{eq:v2a}
\end{equation}
where $\B=2 \nu^2 k_B T/c^2$ is the specific intensity of black-body
radiation in the Rayleigh-Jeans limit. Both $\theta_0$ and$\left(
\frac{\del    B}{\del  T}\right) $  depend on the frequency. In our
analysis we treat these as constants with the value being evaluated at
610 MHz. It is possible to incorporate the effect of the
$\Delta \nu$ dependence of $\theta_0$ and$\left(  \frac{\del
  \B}{\del   T}\right) $ through the  function  $Q(\Delta \nu)$ 
in eq. (\ref{eq:v2a}). This is expected to be a slowly varying function of
$\Delta \nu$ with a variation  of $\sim 1\%$ across the $\Delta \nu$
range of our observation. We have not explicitly considered the function
$Q(\Delta \nu)$ in our present analysis.  This is expected  to introduce
an extra, slowly varying $\Delta \nu$ dependence in the 
estimated $C_{\ell}(\Delta \nu)$. This slowly varying $\Delta \nu$
  dependence, as we shall discuss later, can be included in the
  foreground model and separated from the HI signal which varies
  rapidly with $\Delta \nu$. 
Equation (\ref{eq:v2b}) gives the final expression that we use to
estimate the  angular power spectrum (MAPS)
\begin{equation}
C_{2 \pi U}(\Delta \nu)=87 \ \left(\frac{{\rm mK}}{{\rm Jy}} \right)^2 \ 
\times V_2(\u,\Delta \nu) \,.
\label{eq:v2b}
\end{equation}

In our analysis we have correlated only  pairs of baselines which
satisfy the condition $\mid  \Delta \u \mid  \le 10 \lambda$. We 
have restricted the analysis to baselines $200 \ \lambda \le U \le
5000 \ \lambda$. To test if the visibility correlation is actually 
independent of $\Delta \u$ we have also considered $\mid  \Delta \u
\mid  \le 5 \lambda$ and $20 \lambda$.  The results are unchanged for
$20 \lambda$ and they are rather  noisy for $5 \lambda$, there
being very  few baseline pairs to correlate.

\begin{figure}
\includegraphics[width=90mm,angle=-90]{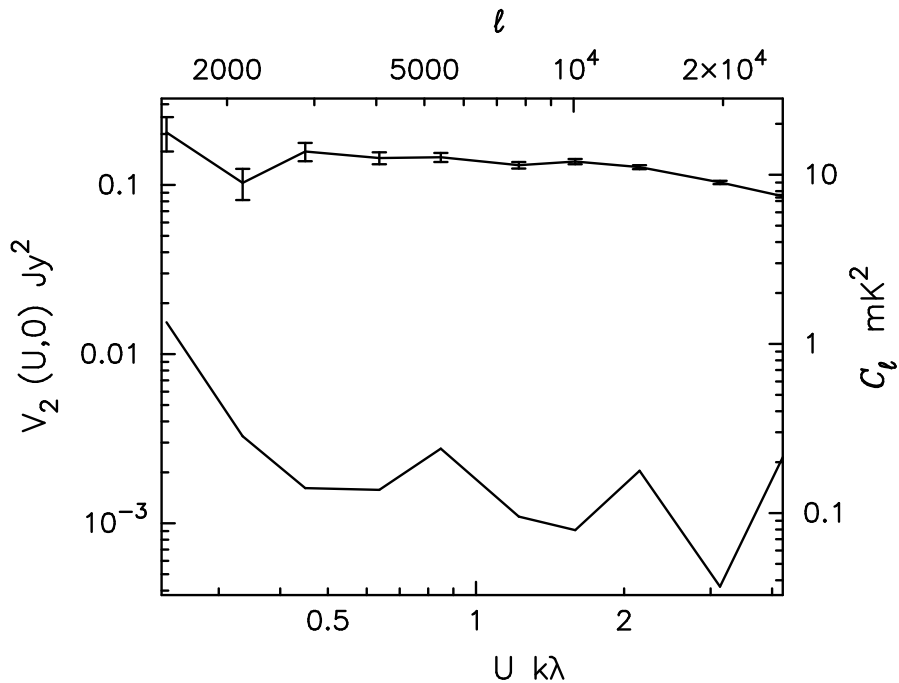}
  \caption{The real (upper) and imaginary (lower) parts of
  the observed visibility correlation $V_{2}(U,0)$. The real part of
  $V_{2}(U,0)$  may be interpreted as 
  $C_{\ell}$ where  $\ell=2 \pi \ U$ (labeled on the right and
  top). The $1-\sigma$ error-bars shown have  contributions from
  both the cosmic variance and system   noise.}  
  \label{fig:vis1}
\end{figure}

The measured $V_2(U,\Delta \nu)$ will, in general, have real and
imaginary parts (Figure \ref{fig:vis1}). As seen in
eq. (\ref{eq:v2b}), the expectation value is predicted to be real, the
expectation value of the imaginary part being zero. We use the real
part of the measured $V_2(U,\Delta \nu)$ to estimate $C_{\ell}(\Delta
\nu)$ through eq.  (\ref{eq:v2b}). A small imaginary part arises due
to the noise in the individual visibilities. This introduces random
fluctuations in both the real and imaginary parts of the measured
$V_2(U,\Delta \nu)$. Figure \ref{fig:vis1} shows the measured
$V_2(\Delta \nu)$ and the inferred $C_{\ell}(\Delta \nu)$ for $\Delta
\nu=0$. As expected, the imaginary part is much smaller than the real
part of $V_2(U,\Delta \nu)$.  Note that we use the notation $C_{\ell}
\equiv C_{\ell}(\Delta \nu=0)$.
 
We next consider the expected statistical fluctuations (error) in
$V_2(U,\Delta \nu)$. 
The total error has two parts i.e., system noise and the cosmic 
variance. The total variance $[\Delta V_2(U_i,\Delta \nu)]^2$ can be
calculated as 
\begin{equation}
[\Delta V_2(U_i,\Delta
  \nu)]^2 \simeq  \frac{\langle N^2 \rangle^2 }{2 N_P}+
  \frac{[{V_2(U_i,\Delta \nu)}]^2}{N_E}
\label{eq:e1}
\end{equation}
 
where $\langle N^2 \rangle =\langle N  N^{*} \rangle $ is the variance
of the noise contribution $N$ in the visibilities that we use in our 
analysis, $N_P$ is the total number of  baseline pairs  that
contribute to $V_2(U,\Delta \nu)$ and  $N_E$ is the number of
independent estimates of $V_{2}(\u,\Delta \nu)$.   Here $\langle N^2
\rangle=\sigma^2$ where  $\sigma$ is the rms. noise, for a   single
polarization,  in the real part (or equivalently the imaginary part)
of a visibility. The value of $\sigma$ is expected to be
(\citealt{thompson})
\begin{equation}
\sigma
=\frac{\sqrt2k_BT_{sys}}{A_{eff}\sqrt{\Delta \nu
      \Delta t}}
\label{eq:rms}
\end{equation}
where $T_{sys}$ is the total system temperature, $k_B$ is the Boltzmann
constant, $A_{eff}$ is the effective collecting area of each
antenna, $\Delta \nu$ is the channel width and $\Delta t$ is correlator
integration time. For the parameters of our observations, $T_{sys}
\approx 100\ {\rm K}$,  $2 T_{sys} k_B/A_{eff}=300 \ {\rm Jy}$, $\Delta
\nu = 0.125 \ {\rm MHz}$ and $\Delta t = 16 \ {\rm s}$ we have
$\sigma^2= 2.25 \times 10^{-2} \ {\rm Jy}^2$. In our analysis we have
used $\langle N^2 \rangle = 1.25 \times 10^{-1} \ {\rm Jy}^2$ which is
the sum of the variance of the real and imaginary components of the
measured visibilities. In our observation  the total error is
dominated by the cosmic variance which is  a few orders of magnitude
larger than the system noise in the entire $U$ range that we have
considered.  

\begin{figure}
\includegraphics[width=150mm,angle=-90]{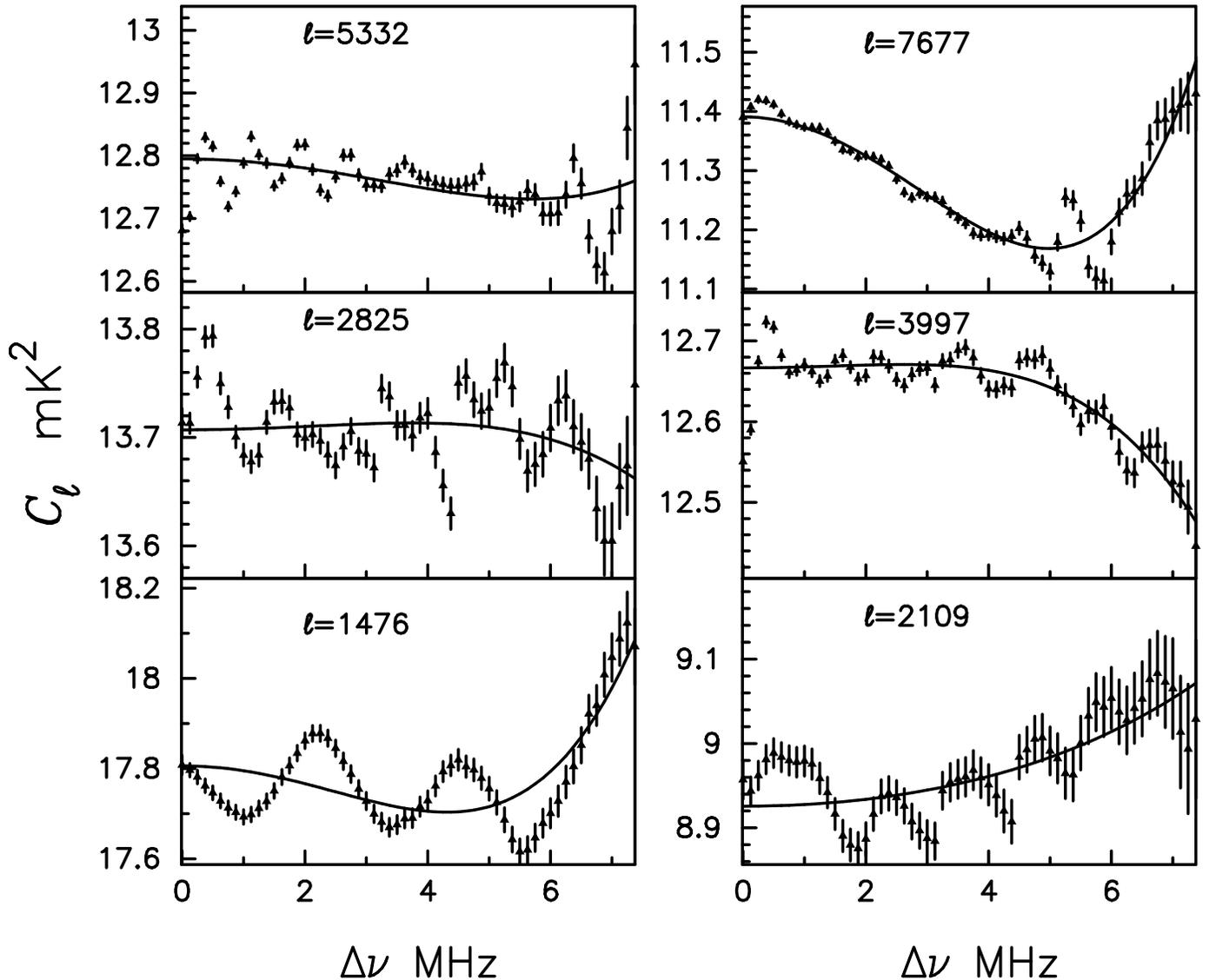}     
\caption{The measured $C_{\ell}(\Delta \nu)$ as a function of $\Delta
  \nu$ for the $\ell$ values shown in each panel. The error-bars
  include only $3 \sigma$ system noise. The solid line shows the $4\,
  \rm th$ order polynomial fits. The details of the polynomial fitting
  procedure are discussed in Section $6$.}
\label{fig:nu1}
\end{figure}

The $\Delta \nu$ dependence of $C_{\ell}(\Delta \nu)$ is shown in
Figure \ref{fig:nu1}. We have considered $U$ values below $1.25 \ {\rm k}
\lambda$ where $\ell = 2 \pi U$.  
As discussed later, the HI
signal   falls at $U > 1 \ {\rm k}\lambda$ which is why we have not 
considered baselines   larger than $1.25 \ {\rm k} \lambda$.
We find that for nearly all the
values of $\ell$ shown in the figure the variation in $C_{\ell}(\Delta
\nu)$ with $\Delta \nu$ is roughly between $0.2 \ {\rm mK}^2$ 
to $0.6 \ {\rm mK}^2$ across the $7.5
\ {\rm MHz}$ band. The fractional variation in $C_{\ell}(\Delta
\nu)$ ranges from $1.5 \ \%$ to $3.6 \ \%$. 
We note that an oscillatory pattern is visible 
in $C_{\ell}(\Delta \nu)$ at nearly all values of $\ell$. The pattern
is most pronounced at the lower $\ell$ values. 
The error-bars shown
in Figure \ref{fig:nu1} include only the system noise
contribution. The measured $C_{\ell}(\Delta \nu)$ is expected to be
dominated by foregrounds which are believed to be largely independent
of $\Delta \nu$. For a fixed $\ell$ the cosmic variance then is
expected to introduce the same error (independent of $\Delta \nu$)
across the entire band. As a consequence we do not consider the cosmic
variance for the $\Delta \nu$ dependence shown in Figure
\ref{fig:nu1}.

The two dimensional (2D) Fourier transform relation between the sky
brightness and the 
visibilities assumed in deriving eq. (\ref{eq:v2a})  is not strictly
valid for GMRT's  field of view  $(\theta_{\rm
  FWHM}=43^{'}$). In addition to $uv$ which are the 
components of the baseline in the plane normal to the direction of
observation, it is also necessary to consider $w$ the component along
the observing direction. To assess the impact of the $w$ term
we have repeated the analysis using only a limited range of baselines
for which $w \le 0.5 \ \times \ U$. We find that limiting the maximum
$w$ value does not make any qualitative change in our results.

\section{The expected redshifted HI  $21$ cm Signal}
Our observing frequency $\nu=610 \ {\rm MHz}$ corresponds to a
redshift  of $z\;=1.32$ for the HI  $21$-cm  radiation. 
Observations of Lyman-$\alpha$ absorption lines seen in quasar spectra  
indicate that the ratio of
the density $\rho_{\rm gas}(z)$ of neutral gas to the present critical
density $\rho_{\rm crit}$ of the universe has a nearly constant value
$\rho_{\rm gas}(z)/\rho_{\rm crit} \sim 10^{-3}$, over a large redshift
range $0 \le z \le 3.5$. This implies that the mean neutral fraction
of the hydrogen gas is $ \bar{x}_{\HI}=50\,\,\Omega_{\rm gas} h^2
(0.02/\Omega_b h^2) =2.45 \times 10^{-2}$  which we adopt for 
our analysis.  The redshifted $21 \, {\rm cm}$ radiation from the HI
will be seen in emission  as a very faint
background in our observation.   The fluctuations in this
background with angle and frequency is a direct probe of the HI
distribution at the redshift $z=1.32$ where the radiation
originated.  We calculate
the MAPS for the redshifted $21$-cm signal  \citep{kanan} using
\begin{equation}
C_l(\Delta \nu)=
\frac{\bar{T}^2~ }{\pi r_{\nu}^2}
\int_{0}^{\infty} {\rm d} k_{\parallel} \, 
\cos (k_{\parallel}\, r'_{\nu}\, \Delta \nu) \, P_{\rm HI}({\bf k}) \,,
\label{eq:sig1} 
\end{equation}
where the three dimensional wave vector ${\bf k}$ has been decomposed
into components $k_{\parallel}$ and $l/r_{\nu}$, along the line of
sight and in the plane of the sky respectively. The comoving distance
$r_{\nu}$ is the distance at which the HI radiation originated. Note
that $(1+z)^{-1} \, r_{\nu}=d_{\rm A}(z)$ is the angular diameter
distance and $r_{\nu}^{'}=d r_{\nu}/d \nu$. The temperature occurring
in eq.~(\ref{eq:sig1}) is given by
\begin{equation}
\bar{T}(z)=4.0 \, {\rm mK}\,\,(1+z)^2  \, \left(\frac{\Omega_b
  h^2}{0.02}\right)  \left(\frac{0.7}{h} \right) \frac{H_0}{H(z)} \,,
\label{eq:sig2}
\end{equation}
and $P_{\rm HI}({\bf k})$ is the three dimensional power spectrum of
the ``21 cm radiation efficiency in redshift space'' \citep{BA5}
which in this
situation is given by
\begin{equation}
P_{\rm HI}(\k)=\bar{x}^2_{\HI} b^2 \left( 1+ \beta  \mu^2 \right)^2
P(k) \,.
\label{eq:sig3}
\end{equation}
The term $\left( 1+ \beta  \mu^2 \right)^2$  arises because of
the HI peculiar velocities \citep{BNS,BA1}, which we assume 
 to be determined by
the dark matter. This is the familiar redshift space distortion seen
in galaxy redshift surveys, where $\beta$ is the linear distortion
parameter and $\mu=k_{\parallel}/k$.  On the large scales of interest
here, it is reasonable to 
assume that HI traces the dark matter with a possible linear bias $b$,
whereby the three dimensional HI power spectrum is $b^2 P(k)$, where
$P(k)$ is the dark matter power spectrum at the redshift where the HI
signal originated. Unless mentioned otherwise, we use the values
$(\Omega_{m0}, \Omega_{\Lambda0},h,\sigma_8,n_s)=(0.3,0.7,0.7,1.0,1.0)$
for the cosmological parameters and $b=1$ for the bias throughout.

\begin{figure}
\includegraphics[width=90mm,angle=-90]{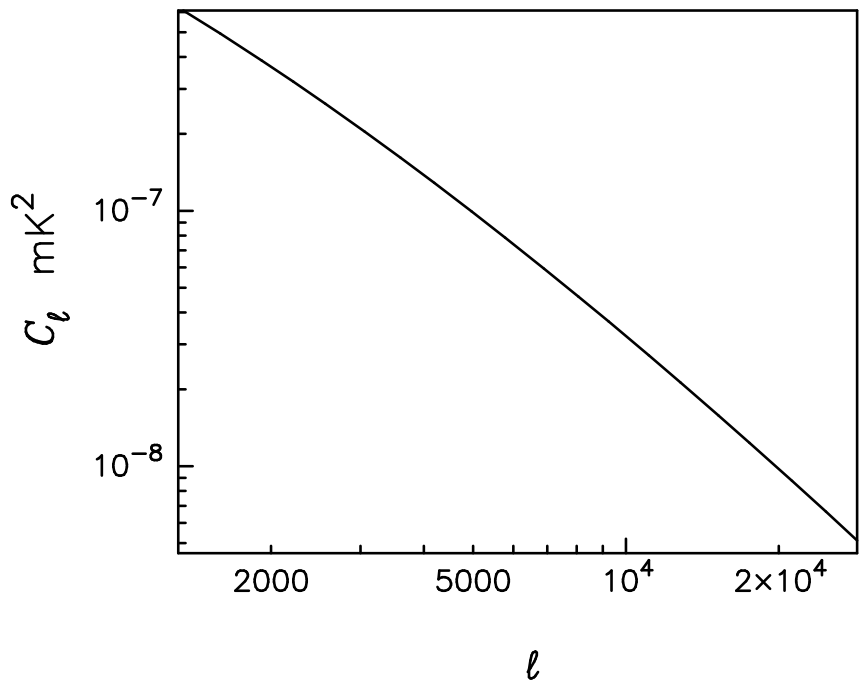}
  \caption{The predicted $C_{\ell}$ for the $610$ MHz
  redshifted 21-cm signal.} 
  \label{fig:s1}
\end{figure}

\begin{figure}
\includegraphics[width=90mm,angle=-90]{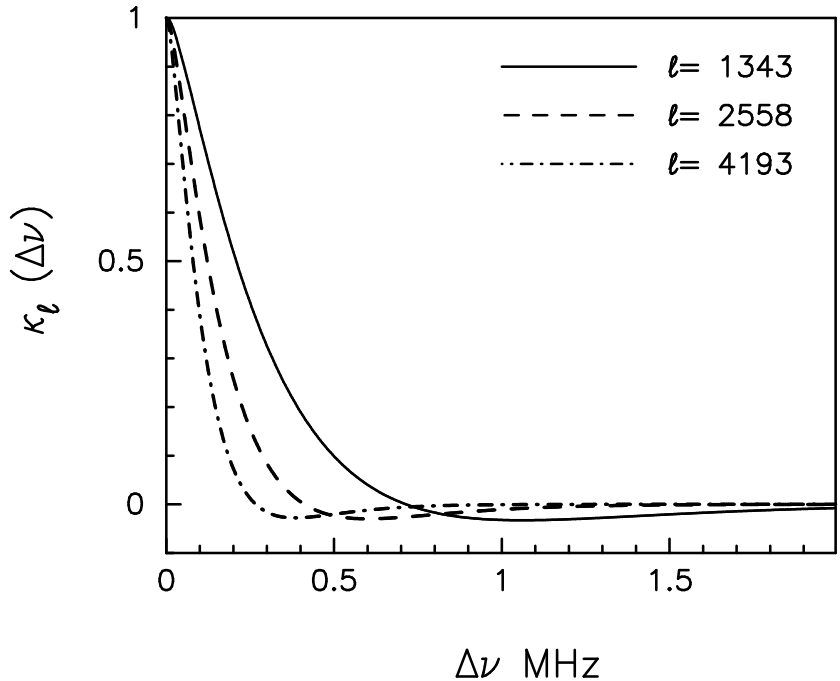}
\caption{The predicted $\kappa_{\ell}(\Delta \nu)$
 for the  $610$ MHz 21-cm signal at three different $\ell$ values
 in the range $1000$ to $5000$.}
\label{fig:s2}
\end{figure}

Figure \ref{fig:s1} shows  $C_{\ell} \equiv C_{\ell}(\Delta \nu=0)$
for the expected HI signal. The $\Delta \nu$ dependence has been
shown (Figure \ref{fig:s2}) through the frequency decorrelation
function $\kappa_{\ell}(\Delta \nu)$    which is defined as
\begin{equation}
\kappa_{\ell}(\Delta \nu)=\frac{C_\ell(\Delta
  \nu)}{C_\ell(0)}\,. 
\end{equation}
This function quantifies  how quickly the HI signal decorrelates
as we increase the frequency separation $\Delta \nu$, with the signal
being correlated and uncorrelated  when $\kappa_{\ell}(\Delta \nu)
\sim 1$ and  
 $\kappa_{\ell}(\Delta \nu)\sim 0$ respectively. \citet{BP} have
proposed an 
analytic formula   that approximates the  visibility correlation
$V_2(U,\Delta \nu)$ for  the GMRT $610$ MHz HI signal. Using this we
obtain  the  analytic expression 
\begin{equation}
C_{\ell}(\Delta \nu)=A \left(\frac{1000}{\ell}\right)^{\gamma} \ 
\exp\left(- \frac{ \Delta \nu}{b_{\ell}} \right)
 {\rm \ sinc} \left(\frac{2 \pi \  \Delta \nu}{d_{\ell}} \right)
\label{eq:sig4}
\end{equation}
with $A=8.0 \times 10^{-7} {\rm mK}^2$, $\gamma=1.2$, $b_{\ell} =0.48
\ \times (1000/\ell)^{0.8} \ {\rm MHz}$ and $d_{\ell}= 1.8 \ \times 
(1000/\ell)^{1.2} \ {\rm MHz}$  which matches the numerically
calculated values of $C_{\ell}(\Delta \nu)$ at around $\sim 10 \%$ for
the $\ell$ and $\Delta \nu$ range of our interest.

We find that the predicted $C_{\ell}$ declines rapidly ($\propto
\ell^{-1.2}$) with increasing $\ell$.  Based on this we have
restricted our analysis to the $\ell$ range $1000$ to $5000$ where the
signal is expected to be strongest. The signal would be larger at
$\ell < 1000$, but the GMRT's field of view restricts us from
accessing these $\ell$ values.  The  measured $C_{\ell}$  values are
around $10^7$ to $10^8$  times larger than the predicted HI signal.  
The predicted signal decorrelates rapidly with increasing $\Delta
\nu$ and  $\kappa_{\ell}(\Delta \nu)$ falls by $90 \%$ or more 
($\kappa_{\ell}(\Delta \nu) < 0.1$) at $\Delta \nu = 0.5 \ {\rm
  MHz}$.   The value of $\Delta \nu$ where $\kappa_{\ell}(\Delta \nu)$ 
falls by $90 \%$ is smaller for  larger values of $\ell$. Further, 
we find that the expected HI signal is  anti-correlated at large values of
$\Delta \nu$ ($\sim 0.8 \ {\rm MHz}$)
where $\kappa_{\ell}(\Delta \nu)$ has a small negative
value. 

\section{Foreground model predictions}

The radiation coming from different astrophysical sources, other than
the HI signal, contribute to the foreground radiation.  Here we mainly
focus on the two most dominant foreground components namely
extragalactic point sources and the diffuse synchrotron radiation from
our own Galaxy.  The free-free emissions from our Galaxy and external
galaxies (\citealt{Shaver}) make much smaller contributions though each of
these  is individually larger than the HI signal. We have modeled the
MAPS for each foreground component as
\begin{equation}
C_{\ell}(\Delta\nu)=A\left(\frac{1000}{\ell}\right)^{\gamma}
\kappa_{\ell}(\Delta\nu) 
\label{eq:fg}
\end{equation}
where $A$, $\gamma$ and $\kappa_{\ell}(\Delta \nu)$ are the amplitude,
  the power law index and frequency decorrelation function
  respectively. The different foreground components  considered here
  are all continuum radiation which are known to  vary smoothly
  with frequency. For each component we denote the spectral index
  using $\alpha$, whereby  the amplitude scales as $A
  \propto \nu^{2 \alpha}$. The value of $A$, whenever used in this
  paper, is at a fixed frequency of $610 \ {\rm MHz}$. The continuum
  nature of the foreground components also implies that  we  expect
  $\kappa_{\ell}(\Delta\nu)$ to be of order unity and 
  vary smoothly with $\Delta \nu$.  The foregrounds  will
  remain correlated    across the frequency   band of our observation,   
 unlike the   HI signal which decorrelates rapidly
 within $\Delta \nu=0.5 \ {\rm MHz}$. Given the absence of any direct
 observational constraints on $\kappa_{\ell}(\Delta \nu)$ for any of
 the foreground components at the angular scales and frequencies of
 our interest, we do not attempt to make any model predictions for
 this quantity beyond assuming that it varies smoothly with $\Delta
 \nu$ across the frequency band of observation. In the subsequent
 discussion we focus on model predictions of   $A$ and $\gamma$ 
which are tabulated in Table~\ref{tab:parm}  for the  
different foreground components. 

Extra-galactic point sources are expected to dominate the sky at
$610 \ \rm MHz$. We have estimated the point source contribution using 
 the  $610 \ \rm MHz$ differential source count from \citet{garn}.
This is the  average differential source count of $610 \ \rm MHz$ GMRT
observations in three different fields of view namely the 
 Spitzer extragalactic First Look, ELAIS-N1 and Lockman Hole surveys. 
 The differential source count in the 
flux range  of their observation ($\sim 0.3 \, \rm to\, 200\ \rm
mJy$) is well fitted by a  single power law 
\begin{equation}
  \frac{dN}{dS}= \frac{1259}{Jy \cdot Sr}\cdot\,\left(\frac{S}{1
  Jy}\right)^{-1.84}\,.
 \label{eq:ds}
\end{equation}
We have assumed that the same power law also holds for  the fainter sources
below the detection limit.  

Point  sources make  two distinct contributions to the angular power
spectrum, the first being the   Poisson noise due to the discrete
nature of the sources and the second arising  from  the  clustering
of the sources.  
The Poisson contribution, which is independent of $\ell$, is
calculated using 
\begin{eqnarray}
C_l =    \left(\frac{\del B}{\del   T}\right)^{-2}
  \int_0^{S_{c}} S^2 \ 
\frac{dN}{dS} \  dS  \,,
\label{eq:ps}
\end{eqnarray}
where $S_{c}=250 \ {\rm m Jy}$  is the flux of the brightest source
in our field of 
view. The uncertainty in the Poisson contribution  involves the
fourth moment $  \int_0^{S_{c}} S^4 \ \frac{dN}{dS} \  dS  $ of the
source count and is given by 
\begin{equation}
[\Delta C_{\ell}(0)]^2=\left(\frac{S_c}{{\rm
    Jy}}\right)^{2.32}\left[\,\, 69.63- 
 133.15\,\left(\frac{S_c}{{\rm Jy}}\right)^{0.84}\,\, \right] \,.
\end{equation} 

The analysis of large samples of nearby radio-galaxies has shown that
the point sources are clustered. \citealt{cress} have measured the
angular two point correlation function at $1.4 \, \rm GHz$ (FIRST
Radio Survey) , across an angular scale of $.02^{o}\, \rm to\, 2^{o}$,
equivalent to a $\ell$ range of $90<\ell<9000$. Throughout the entire
angular scale the measured two point correlation function can be well
fitted with a single power law of the form
$w({\theta})=(\theta/\theta_{0})^{-\beta}$ , where $\beta = 1.1$ and
$\theta_{0} = 17.4'$. This partly covers the range of angular scale
($\sim 10^{'}$ to $20^{''}$ or $\sim 1000<\ell<3\times10^4$) that we
are interested in. We will assume that the clustering of the sources
remain unchanged at our observing frequency.  They have also reported
that on small scales ($< 0.2^{o}$) the double and multicomponent
sources tend to have a larger clustering amplitude than that of the
whole sample. They also found that the sources with flux densities
below $2\,\rm mJy$ have a much shallower slope ($\sim
0.97$) for the measured correlation function. It seems that the
amplitude and slope of the measured two point correlation function
changes with the angular scale and flux densities of the sources. For
our present purpose we have used
$w({\theta})=(\theta/\theta_{0})^{-1.1}$ which have been measured up to
$\ell= 9000$. We have assumed that the slope of the two point
correlation function will remain unchanged beyond  $\ell= 9000$ . We
then have

\begin{eqnarray}
C_{\ell} = \left(\frac{\del B}{\del   T}\right)^{-2} 
\left( \int_0^{S_{c}} S \
\frac{dN}{dS} \  dS  \right)^2 \  w_{\ell}
\label{eq:clus}
\end{eqnarray}
where  $ w_{\ell} \propto {\ell}^{\beta -2}$ is the angular power
spectrum which is the Fourier transform of $w(\theta)$.

The Galactic diffuse synchrotron radiation is believed to be produced
by cosmic ray electrons propagating in the magnetic field of the
Galaxy \citep{GS69}. The angular power spectrum is predicted  to scale as
$\ell^{-\gamma}$ with $\gamma \approx 2.4$ \citep{teg00} to angular
scales as small as $4^{'}$ , and the spectral index has a value $\sim
2.8$. Here we have extrapolated the parameters from the 
$130 \ {\rm MHz}$ model prediction of 
\citet{Santos}. Recently \citet{bernardi} have characterized the
  power spectrum of the total diffuse radiation at $150 \, {\rm MHz}$
  at the angular scales of our interest. The $\ell$ dependence  that
  we adopt   in our foreground model is consistent with that found
  by   \citet{bernardi}.

The total error in our  model predictions is calculated
by adding the variances from different contributions.

\begin{figure}
\begin{centering} 
\includegraphics[width=90mm,angle=-90]{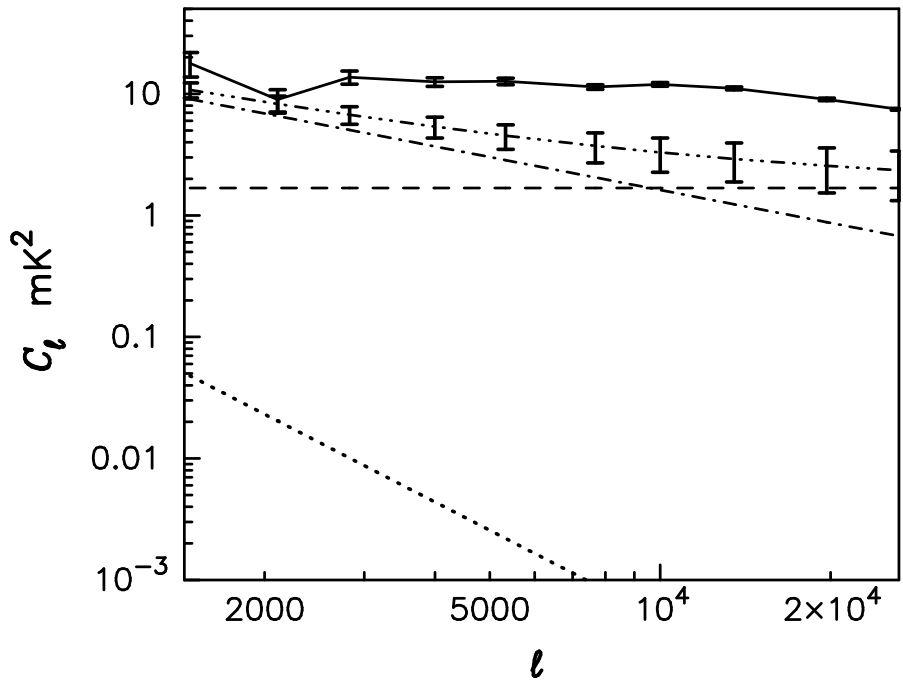}
\caption{The black solid line shows the observed $C_\ell$ 
with $1-\sigma$ error-bars, whereas the dash-dot-dot-dot curve shows
the total foreground model prediction with its $1-\sigma$ error-bars. 
The  dashed and the dot-dashed  curves are respectively the Poisson
and clustering contributions from the  point sources, while the lower
most dotted curve is the contribution from  Galactic  synchrotron
radiation .} 
\label{fig:fore_clus} 
  \end{centering}
\end{figure}

Figure \ref{fig:fore_clus} shows the point source and synchrotron
contributions along with the total measured signal. At large angular
scales $(\ell \le 10^4)$ the foreground model prediction is dominated
by the clustering of point sources, the point source Poisson
contribution being the second largest component at these angular
scales. This is reversed at smaller angular scales $(\ell > 10^4)$
where the point source Poisson contribution dominates and the
clustering component is the second largest contribution. The Galactic
synchrotron contribution, also shown in Figure  \ref{fig:fore_clus},
is much smaller at all the angular scales of 
our interest. 
The contributions from Galactic and extra-galactic free-free emission,
whose parameters have been extrapolated from \citet{Santos}, are also
listed in Table~\ref{tab:parm}. These are much smaller and hence are
not shown in  Figure \ref{fig:fore_clus}.
 The expected HI signal $(C_{\ell} \sim 10^{-6} - 10^{-7}
\ {\rm m K}^2)$ is much smaller than all the foreground components
mentioned here, and is not shown in the figure.

We find that the measured $C_{\ell}$ is within the $1-\sigma$
error-bars of the model prediction, for $\ell \le 2300$. The measured
$C_{\ell}$ is around three times larger than the model predictions at
smaller angular scales where the measured values do not lie within the
$1-\sigma$ error-bars of the model predictions. The source of this
discrepancy is, at present, unknown to us. The model predictions
require the source properties to be extrapolated to faint flux levels
and small angular scales where direct observations are not
available. It is possible that the model predictions have been
underestimated.  For the present work we assume that the measured
$C_{\ell}$ is correct and that the model predictions have been
underestimated at small angular scales.


For the subsequent analysis
in this paper we shall assume that the measured $C_{\ell}(\Delta \nu)$
is a combination of contributions from foregrounds, the HI signal and
noise. Further, the HI signal being several orders of magnitude
smaller than the foregrounds, we may interpret the measured
$C_{\ell}(\Delta \nu)$ as an estimate of the foregrounds {\it actually
present} in our field of view.

\begin{table}
\caption{Values of the parameters used for characterizing
  different foreground contributions at 610 MHz. Here $S_c$ is the
  flux of the brightest source in the field of view.}
\vspace{.2in}
\label{tab:parm}
\begin{tabular}{|p{4cm}|p{3cm}|p{1cm}|p{1cm}|}
\hline
Foregrounds&$A({\rm mK^2})$&${\alpha}$&${\gamma}$\\ 
\hline
Point source&$20.03\times\left (\frac{S_{cut}}{\rm Jy}\right)^{0.32}$&$2.07$&$0.9$\\
(Clustered part)& & \\
\hline
Point source&$8.38\times\left(\frac{S_{cut}}{\rm Jy}\right )^{1.16}$&$2.07$&$0$\\
(Poisson part)& & \\
\hline
Galactic synchrotron&$0.122$&$2.80$&$2.4$\\
\hline
Galactic free-free&$1.14\times10^{-4}$&$2.15$&$3.0$\\
\hline
Extra Galactic free-free&$2.11\times10^{-5}$&$2.1$&$1.0$\\
\hline

\end{tabular}
\end{table}

\section{Foreground removal}
\label{sec:fgrem}
Removing the foregrounds which, as we have seen, are several orders of
magnitude larger than the HI signal is possibly the biggest challenge
for detecting the HI signal. There have  been quite a few earlier
works on this, nearly all either theoretical or simulation. 
All attempts in this direction are based on the assumption that the
foregrounds are continuum radiation which vary slowly with frequency
whereas the HI is a line emission which  varies rapidly with
frequency.  

A possible line of approach is to represent the sky signal as 
an image cube where in addition
to the two  angular coordinates on the sky we have the frequency as
the third dimension. For each  angular position, polynomial fitting is
used to subtract out the component of the sky signal that varies
slowly with 
frequency. The residual sky signal is expected to contain only the HI
signal and noise \citep*{jelic,bowman,Liu1}.  \citet{liu} show that
this method of foreground removal has problems which could be
particularly severe at large baselines if the $uv$ sampling 
is sparse. They 
propose an alternate method where the frequency dependence of the
visibility data is fitted with a  polynomial and this is used to
subtract out the slowly varying component. The residuals are expected
to contain only  noise and the HI signal.

In this work we have
attempted to subtract out the brightest point sources from the image 
using standard AIPS tasks. We have used the AIPS task 'UVSUB' to
subtract the Clean Components (CC) of the brightest sources from the
visibility data. Continuum images  were used for this purpose. 
The resulting visibility data was used to make a new image. We find 
that  this method fails to remove the point sources efficiently,
several imaging artifacts  remain in the vicinity  of
 bright sources even after the sources have been
removed. Similar findings were reported in \citet{ali} where the same
technique was used  to remove point sources from $150 \ {\rm MHz}$
GMRT observations. Given the poor performance of this image based
technique, we have not pursued it any further. The visibility based
technique proposed by  \citet{liu} requires the data to be gridded in
 $uv$ plane. The estimator that we have used to determine
$C_{\ell}(\Delta \nu)$ (Section \ref{sec3}) works with  the individual
visibilities. Using the gridded data would introduce a positive noise
bias in $C_{\ell}(\Delta \nu)$ and hence we do not adopt this
technique here.

The foreground subtraction techniques discussed above  all attempt to
 remove the foregrounds {\it before} determining the angular power
 spectrum. Here we propose a different method where the foregrounds
 are subtracted {\it after} determining the angular power
 spectrum. 
The measured    $C_{\ell}(\Delta \nu)$ (Figures
 \ref{fig:vis1} and \ref{fig:nu1}) 
is  a sum of the foregrounds, noise and the HI signal. The HI signal
 decays rapidly with increasing $\Delta \nu$.  This contribution is
less than $10 \, \%$ for  $\Delta \nu \ \ge 0.5 \, {\rm MHz}$ and it is 
negligibly small  for $\Delta \nu \ > 1 \, {\rm MHz}$ (Figure
 \ref{fig:s2}). We  assume that $C_{\ell}(\Delta \nu)$
 measured in  the  frequency interval $0.5 \ {\rm MHz}  \le \Delta \nu
 \le \ 7.5 \ {\rm  MHz}$   contains only    foreground and noise. 
 Further, we assume that 
 the foreground contribution to $C_{\ell}(\Delta \nu)$
 has a
 slow  $\Delta \nu$ dependence  which can be well fitted by a low
 order  polynomial. Note that,  in addition
 to the intrinsic  $\Delta \nu$ dependence of the foreground, the
 measured $C_{\ell}(\Delta \nu)$  has an additional   $\Delta \nu$
 dependence arising from the factor $Q(\Delta \nu)$
 (eq. \ref{eq:v2a}). The latter is a slow, monotonic variation and we
 expect that both these effects can be adequately accounted for by 
a  low order polynomial. 
 We use the interval  $0.5 \ {\rm MHz} \le  
 \Delta \nu  \le \ 7.5 \ {\rm  MHz}$ to estimate this polynomial,
 which is then 
 used  to subtract   the foreground   contribution  from $C_{\ell}(
 \Delta \nu)$  across the entire  range of our  measurement ($0 \le
 \Delta \nu \le 7.5 \  {\rm MHz}$).   The residual $C_{\ell}(\Delta
 \nu)$ is expected to be a sum of only the HI signal and noise.    

In order to  illustrate our technique of foreground subtraction 
and to demonstrate its efficacy,  we first apply it to simulated data
where a known HI signal has been put in by hand. Given the uncertainty
in our current understanding of the foreground properties and of the
effects that have possibly been introduced during the observation  and
the subsequent analysis, we are guided by the measured
$C_{\ell}(\Delta \nu)$ for our simulations. We find that the measured
$C_{\ell}(\Delta \nu)$ (Figure \ref{fig:nu1})  has a value around
$\sim 10 \ {\rm mK}^2$,   with  $\sim 5 \%$ variation with $\Delta
\nu$ across the $7.5 \ {\rm MHz}$ band. Further, the error  has a
typical value $\sqrt{[\Delta C_{\ell}(\Delta \nu)]^2} \ \sim \ 0.01 \
{\rm mK}^2$ (system noise only). We have simulated the measured
MAPS using 
\begin{equation}
C_{\ell}(\Delta \nu)=\sum_{n} a_n \ (\Delta \nu)^n \ + \ \delta + 
\alpha \ C^{\rm HI}_{\ell}(\Delta \nu)
\end{equation}
where the polynomial $\sum_{n} a_n \ (\Delta \nu)^n$ represents the
slowly varying $\Delta \nu$ dependence which causes $C_{\ell}(\Delta
\nu)$ to vary by $\sim 10\ \%$ across the $7.5 \ {\rm MHz}$ band. Our
$C_{\ell}(\Delta \nu)$ estimator (eq. \ref{eq:v2b}) is even in $\Delta
\nu$, and hence we have only considered polynomials of even order. Our
simulation was restricted to fourth order polynomials where the
coefficients $a_0,a_2, a_4$ are Gaussian random variables with mean
$12,0,0 \ {\rm mK}^2$ and rms. $1,10^{-2},10^{-4} \ {\rm mK}^2$
respectively.  The term $\delta$ is a Gaussian random variable of rms.
$0.01 \ {\rm mK}^2$ which incorporates the error and $C^{\rm
HI}_{\ell}(\Delta \nu)$ is the HI signal (eq. \ref{eq:sig1}) . The
noise in our observation is considerably larger than the HI signal and
it would not be possible to detect the signal even if the foregrounds
were perfectly subtracted. The factor $\alpha$ in our simulations
amplifies the HI signal so that it lies above the noise.  The value of
$\alpha$ has been chosen such that $C_{\ell}(\Delta \nu)= 5 \times
0.01 \ {\rm mK}^2$ (5-sigma) at the value of $\Delta \nu$ where
$C_{\ell}(\Delta \nu)$ is $70 \ \%$ of the peak value $C_{\ell}(0)$.
The simulations have exactly the same frequency bandwidth and channel
width as the measured $C_{\ell}(\Delta \nu)$.  Though in this paper we
have only considered fourth order polynomials for our simulations, the
same procedure can easily be repeated considering even polynomials of
any order.

\begin{figure}
 \begin{centering} 
   \includegraphics[width=150mm,angle=-90]{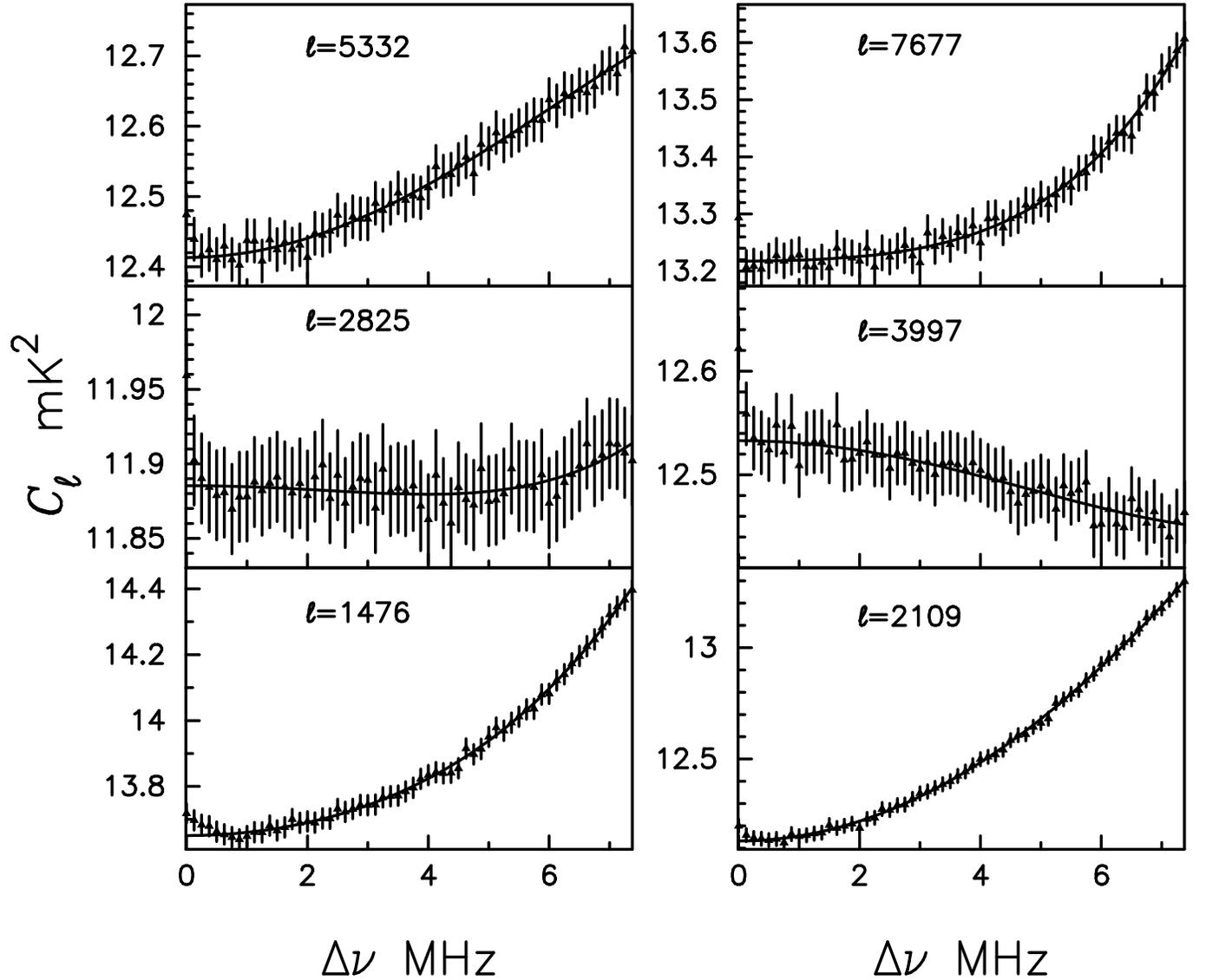}
  \caption{The simulated $C_{\ell}(\Delta \nu)$ with $3\sigma$
    error-bars (system noise) is shown for different values of
    $\ell$. The solid curve shows the best fit $4 \, {\rm th}$ order
    polynomial determined using the interval $0.5 \ {\rm MHz} \le
    \Delta \nu \le 7.5 \ {\rm MHz}$.  }
  \label{fig:sim1} 
  \end{centering}
\end{figure}

\begin{figure}
 \begin{centering} 
  \includegraphics[width=150mm,angle=-90]{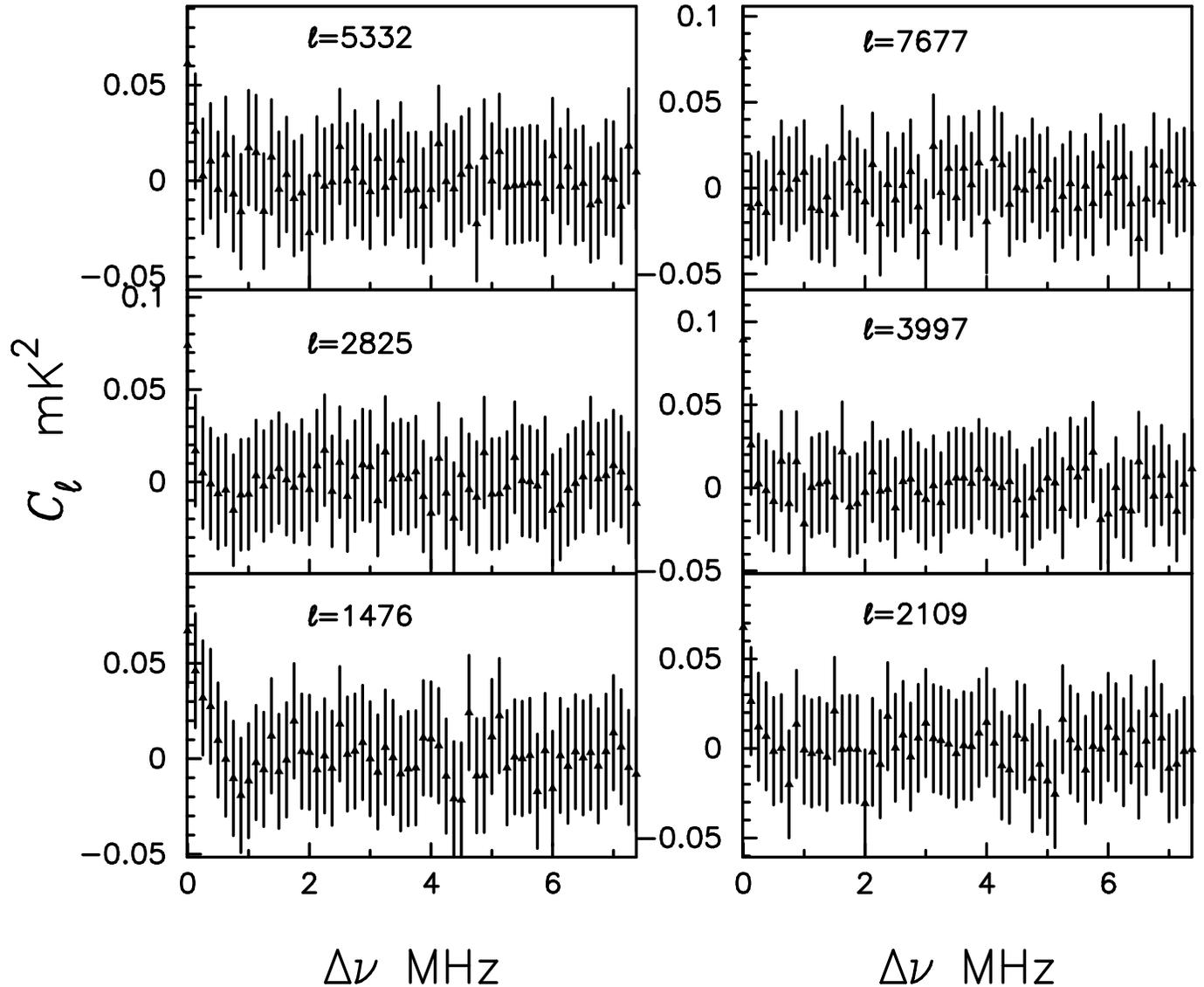}
  \caption{The residual, with $3\sigma$ error bars, after subtracting
    the best fit $4 \, {\rm th}$ order polynomial from the simulated
    $C_{\ell}(\Delta \nu)$.}
  \label{fig:sim2} 
  \end{centering}
\end{figure}

\begin{figure}
 \begin{centering} 
  \includegraphics[width=150mm,angle=-90]{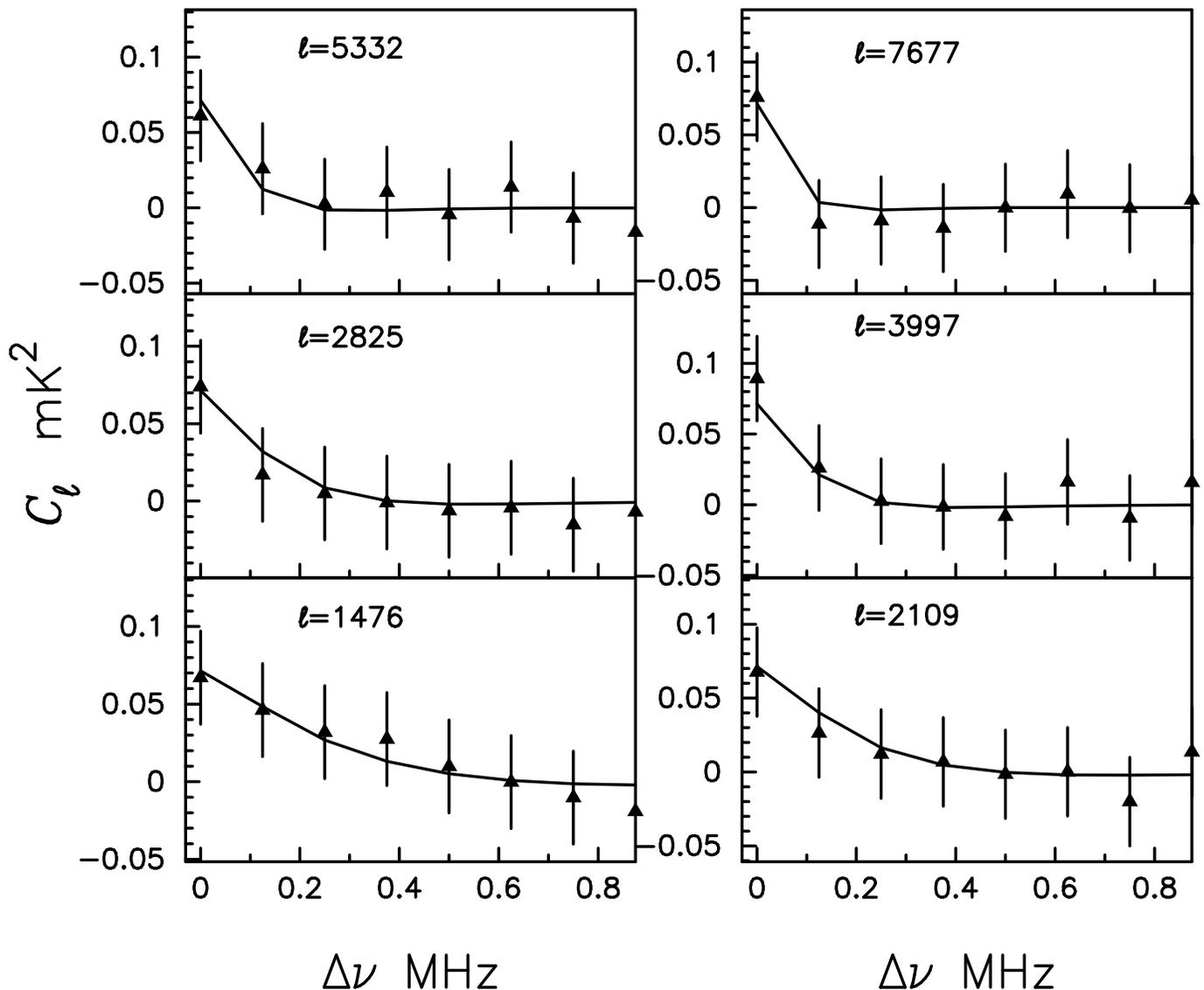}
  \caption{The residual, with $3\sigma$ error bars, after subtracting
    the best fit polynomial from the simulated $C_{\ell}(\Delta
    \nu)$. The solid curve shows the HI signal that had been put in by
    hand in the simulations.  }
  \label{fig:sim3} 
  \end{centering}
\end{figure}

Figure \ref{fig:sim1} shows the simulated $C_{\ell}(\Delta \nu)$  for
the different values of $\ell$. Note that the polynomial coefficients
$a_n$ are different  for each realization
of the simulation. We have fitted the simulated data with a fourth
order polynomial   using the interval $0.5 \ {\rm MHz} \le \Delta \nu \le
7.5 \ {\rm MHz}$. The best fit polynomial is  also shown in  Figure
\ref{fig:sim1}.  The residuals, after the best fit polynomial is
subtracted from the simulated  $C_{\ell}(\Delta \nu)$, are shown in
Figure \ref{fig:sim2}. In the interval  $0.5 \ {\rm MHz} \le
\Delta \nu \le 7.5 \ {\rm MHz}$, the residuals are within $\pm 3
\sigma$ from zero  which is consistent with noise. Figure
\ref{fig:sim3} shows the residuals in the range   $\Delta \nu \le 1 \
    {\rm MHz}$ 
overlaid with the HI signal that had been added by hand.  We find that
our foreground subtraction technique successfully extracts the HI
signal that had been added in the simulated  data, despite its being
buried in foregrounds which are $\sim 200$ times larger.  We note that
we have also tried a slightly  different technique of foreground
subtraction where we have used the entire $\Delta \nu$ range ($\le 7.5
\ {\rm MHz}$) to estimate the polynomial. We find that the latter
technique does not correctly recover the HI signal that had been put
in by hand.

\section{Result and Conclusions}
We have measured the 
statistical properties of the background radiation  across  angular
scales  $20''$ to $10'$  using the 
multi-frequency angular power spectrum $C_{\ell}(\Delta\nu)$.
Frequency channels $20$ to $80$ were used for the
analysis. This corresponds to a total bandwidth of $7.5 \, {\rm MHz}$
with a resolution of $125 \, {\rm kHz}$. The measured
$C_{\ell}(\Delta\nu)$ has values around $12 \,{\rm
  mK^2}$. Considering first the $\ell$ dependence of $C_{\ell}$
(Figure \ref{fig:vis1}),
starting from $\sim 18 \,{\rm mK^2}$ at $\ell \sim 1000$, it drops
to $\sim 9 \, {\rm mK^2}$ at $\ell \sim 2000 $ and then  rise to a
nearly constant value of around $13 \, {\rm mK^2}$. 
 The uncertainty in 
$C_{\ell}$ is mainly due to the sample variance {\it ie.} the fact that we
 have observed a 
 single $\sim 1.5^{\circ} \times \sim 1.5^{\circ}$ field of view which
 gives  a limited number of independent estimates of $C_{\ell}$, the
 system noise  makes a relatively smaller contribution. 
We next consider the $\Delta\nu$ dependence of $C_{\ell}(\Delta\nu)$
for different values of $\ell$ (Figure \ref{fig:nu1}). 
Assuming that the foreground contributions all have a smooth power law
$\nu$ dependence, the expected $\Delta \nu$ dependence may be
estimated 
through a Taylor series expansion as $C_{\ell}(\Delta\nu)=C_{\ell} \,
\left[1 \, +   \, B \left(\frac{\Delta \nu}{\nu}\right)^2 ...\right]$
where $B$ is constant of order unity.
 The odd powers of ${\Delta\nu}/{\nu}$ cancel out because the
 estimator averages positive and negative $\Delta\nu$ values.
 The expected change in $C_{\ell}(\Delta\nu)$ is $\sim 1.5 \times
 10^{-2} \, \%$ for $\Delta\nu = 7.5 \, {\rm   MHz}$. The
measured $C_{\ell}(\Delta\nu)$  (Figure
\ref{fig:nu1}) has a smooth  variation of the order of a
few percent ($1\%$ to $4\%$)  across the $7.5 \, {\rm MHz}$ bandwidth
of  our observation. In addition to the smooth $\Delta
\nu$ dependence, we also notice a small oscillatory pattern in the
measured 
$C_{\ell}(\Delta\nu)$.  The expected HI contribution to
$C_{\ell}(\Delta \nu)$ is $\sim 10^{-7}$ times smaller than the
measured values, and we interpret the measured $C_{\ell}(\Delta \nu)$
as being nearly entirely foregrounds and noise. 

We next consider results for foreground removal using the technique 
discussed in Section \ref{sec:fgrem}. For a fixed $\ell$, 
the frequency range   $0.5 \ {\rm MHz} \le  \Delta \nu \le \ 7.5 \ {\rm
  MHz}$  was used  to estimate a fourth order polynomial fit to 
$C_{\ell}(\Delta\nu)$. The $C_{\ell}(-\Delta \nu)=C_{\ell}(\Delta
\nu)$ symmetry of the $C_{\ell}(\Delta \nu)$ estimator   was applied  in 
the fitting procedure.  This fit was used to subtract out the
foreground contribution from the entire frequency range  
$\Delta \nu \le \ 7.5 \ {\rm  MHz}$.  The performance  of this 
foreground removal technique was assessed by visually inspecting the
fit and the residuals across the entire band.
We find that increasing the order of the polynomial does not
result in any significant improvement, and hence we restrict our
analysis to a fourth order polynomial for which the fits have been
shown in Figure \ref{fig:nu1}. 
\begin{figure}
 \begin{centering} 
  \includegraphics[width=150mm,angle=-90]{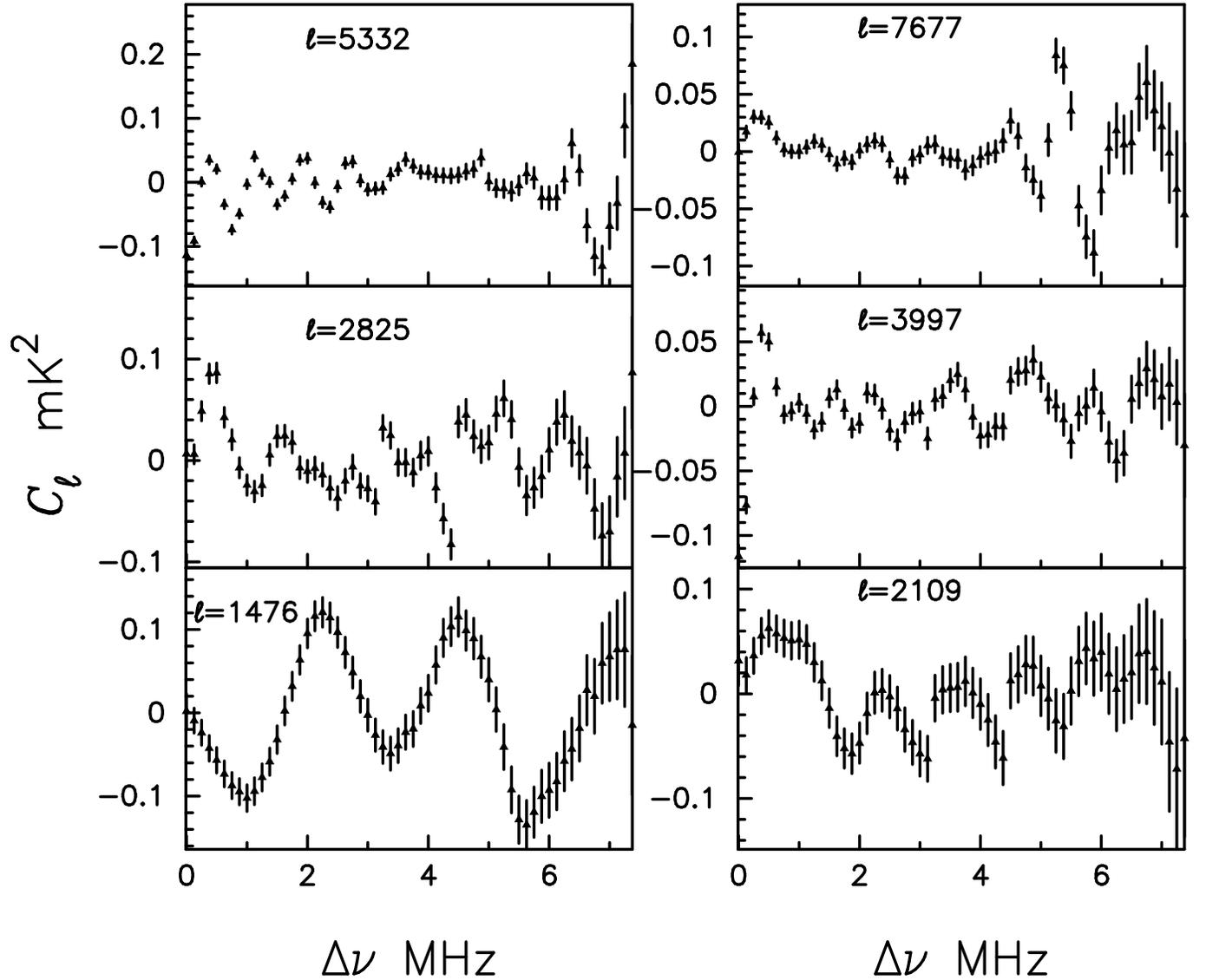}
  \caption{The residual, with $3\sigma$ error bars (system noise
    only), after subtracting the best fit $4\,\rm th$ order polynomial
    from the measured $C_{\ell}(\Delta \nu)$.}
  \label{fig:res1} 
  \end{centering}
\end{figure}
\begin{figure}
 \begin{centering} 
  \includegraphics[width=150mm,angle=-90]{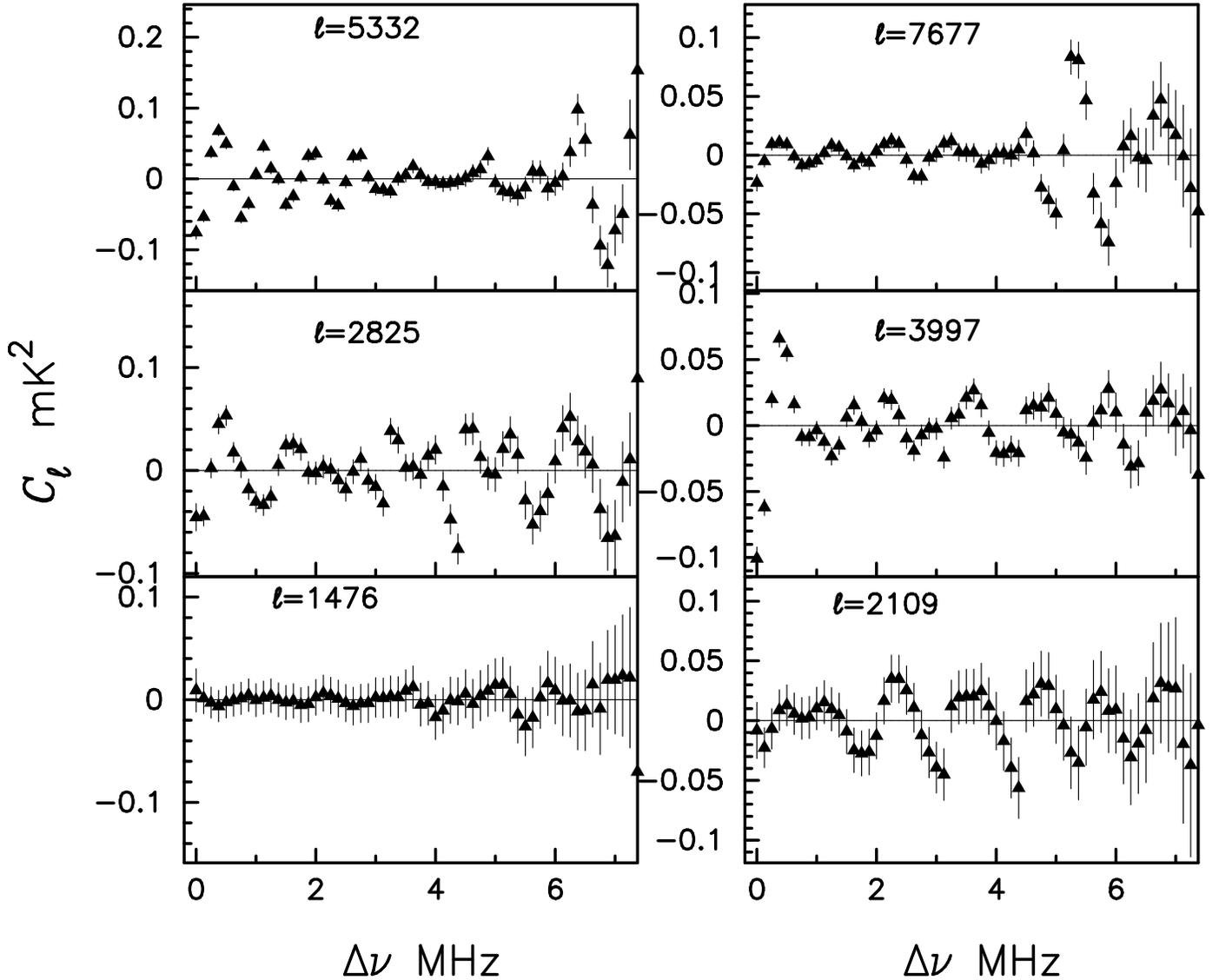}
  \caption{The residuals with $3\sigma$ error  bars (system noise only)
after applying the filter using $m_c=7$.}
  \label{fig:fil} 
  \end{centering}
\end{figure}
\begin{figure}
 \begin{centering} 
  \includegraphics[width=90mm,angle=-90]{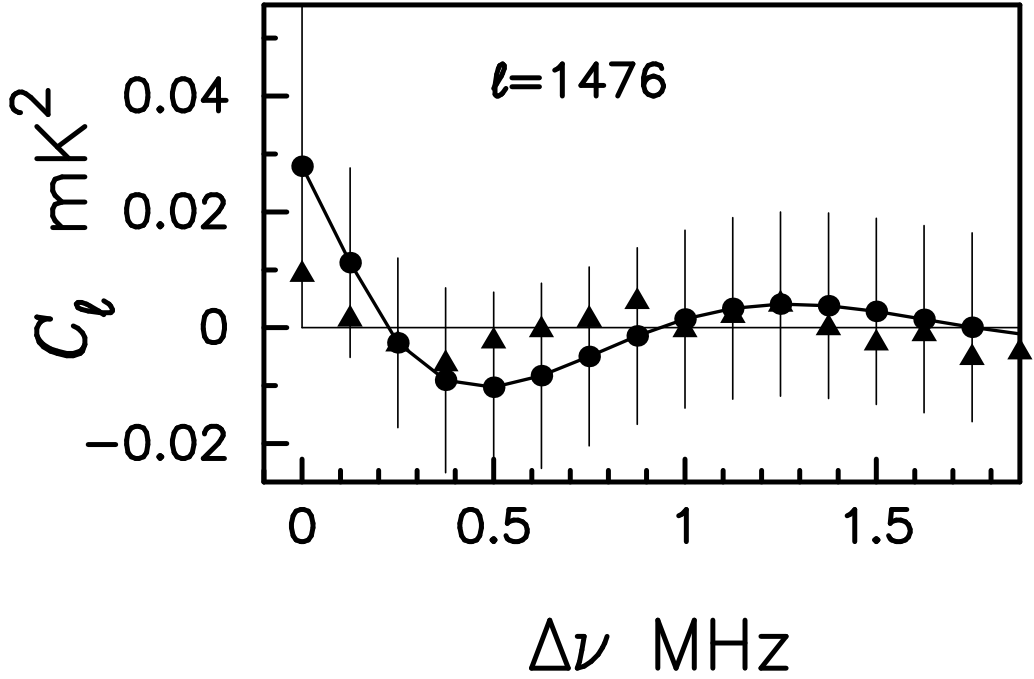}
  \caption{The expected HI signal for $\xb b=7.95$ (circle
  and solid   curve )and  $3\sigma$ error bars (total errors)
after applying the filter with $m_c=7$.  The
  residuals from   the measured $C_{\ell}(\Delta \nu)$ are also shown
  (triangles). } 
  \label{fig:xh1b} 
 \end{centering}
\end{figure}
The residuals in $C_{\ell}(\Delta \nu)$, we find, typically have
values within $0.1 \ {\rm mK^2}$ (Figure \ref{fig:res1}). 
 In all cases the residuals are not consistent with
$C_{\ell}(\Delta \nu)=0$  ({\it ie.} noise only). The residuals, we
 find, have a nearly sinusoidal oscillatory pattern. These
 oscillations are most pronounced for the lowest $\ell$ value where it
 has an amplitude of $\sim 0.1 \ {\rm mK}^2$ and a period of $\Delta
 \nu \sim 3 \, {\rm MHz}$. The period and amplitude both decrease 
 with increasing  $\ell$. The oscillations are possibly not very well
 resolved at the larger  $\ell$ values due to the  $0.125 \, {\rm
   MHz}$ channel width. The oscillations would possibly be more
 distinctly visible in observations with higher frequency resolution. 
The oscillatory residual pattern is quite distinct from the expected
HI signal and also from random noise, and in principle it should be
possible to distinguish between these by considering the Fourier
transform 
\begin{equation}
\tilde{C}_{\ell}(\tau_m)=\sum_n e^{i 2 \pi \,   \tau_m \,  \Delta
  \nu_n} \ C_{\ell}(\Delta \nu_n)
\end{equation}
where $n,m=-59, -58, ...,0,.., 58, 59$, $\Delta \nu_n= n \times 0.125
\ {\rm MHz}$ and $\tau_m=m \, (119 \times 0.125 \ {\rm
  MHz})^{-1}$.  We expect the oscillatory pattern to manifest itself
as a localized feature in $\tilde{C}_{\ell}(\tau)$ and it should be
possible to remove the oscillatory feature by applying a suitable
filter to $\tilde{C}_{\ell}(\tau)$. We find that 
for the smallest $\ell$ values  the amplitude of
$\tilde{C}_{\ell}(\tau)$ is peaked at a few $\tau_m$ values located
within $\mid m \mid \le 10$.  Based on this we have chosen a filter 
\begin{eqnarray}
\tilde{F}(\tau_m)&=&0 \hspace{4cm} \mid m \mid \le m_c \\ 
&=& 1.0-e^{-(\mid m \mid -m_c)^2/2} \hspace{1cm} \mid m \mid > m_c
\nonumber  
\end{eqnarray}
such that $\tilde{F}(\tau_m) \tilde{C}_{\ell}(\tau_m)$ removes the Fourier
components within $\mid m \mid \le m_c$ from the residual
$\tilde{C}_{\ell}(\tau_m)$. Calculating $C_{\ell}(\Delta \nu)$ after
applying the filter, we find that for the smallest $\ell$ value  
the oscillatory pattern is removed
 if we use $m_c=7$ or larger.  The oscillatory pattern is somewhat
 reduced for the next two $\ell$ values while the three largest $\ell$
 values are not much affected by  the filter with $m_c=7$.
It is possible to remove the oscillatory pattern from the second
largest $\ell$ value by increasing the value of $m_c$ to  $m_c=14$,
but the oscillatory pattern still persists for the larger $\ell$
values. Increasing $m_c$ will also reduce the HI signal, and hence we
do not consider $m_c=14$ in the subsequent discussion.  The filter is
also expected to affect the noise estimates, and the noise in the
different $C_{\ell}(\Delta \nu)$ will be correlated as a consequence
of the filter. For $m_c=7$, we are filtering out $\sim 10 \%$ of the
$\tilde{C}_{\ell}(\tau)$ values, and hence we do not expect this to be
a very severe effect . Thus, for the purpose of this paper, it is
reasonable to assume that the noise is unaffected by the filter. 

We find that for the smallest $\ell$ value ($\ell=1476$) the residuals
are consistent with zero at the $3\sigma$ level. Based on this we
conclude that we have successfully removed the foreground contribution
from the measured $C_{\ell}(\Delta \nu)$ at this value of $\ell$. 
The residual oscillatory pattern persists at all the larger $\ell$
values where we are not successful in completely removing the
foregrounds.  The cause  of this oscillatory residual, which at the moment
is unknown to us, is an important issue which we plan to investigate  in
future.   

We next use the measured $C_{\ell}(\Delta \nu)$ at $\ell=1476$  to
place an  
upper limit on  the HI signal. The amplitude of the expected HI signal
is determined by the factor $(\xb \, b)^2$ (eqs. \ref{eq:sig1}
and \ref{eq:sig3}) where $\xb$ and $b$ are the HI neutral
fraction and the HI  bias parameter respectively. In the discussion
till now we have used $\xb  b =2.45 \times 10^{-2}$
to estimate the expected HI signal $C^{\rm HI}_{\ell}(\Delta \nu)$. 
We now  consider $\xb b$ as a free  parameter whose value is
unknown, and  ask if  it is possible to use our observation to
place an upper limit on the value of  $\xb \, b$.
Considering $\xb 
\, b$ as an unknown parameter, the expected HI signal  $ C^{\rm 
  HI}_{\ell}[\xb b](\Delta \nu)$ can be expressed   as  
\begin{equation}
C^{\rm HI}_{\ell}[\xb b](\Delta \nu)=\left[\frac{\xb
  b}{2.45 \times   10^{-2}} \right]^2  C^{\rm HI}_{\ell}(\Delta \nu) 
\end{equation}
The HI signal would be detectable  in our
observation at the $3\sigma$ level if  
\begin{equation}
C^{\rm HI}_{\ell}[\xb b](\Delta \nu) > 3 \sqrt{ \{C^{\rm
    HI}_{\ell}[\xb b](\Delta \nu)\}^2/N_E + \{\Delta
    C_{\ell}(\Delta \nu)\}^2_{sys} }
\label{eq:det}
\end{equation}
where $N_E$ is the number of independent estimates of the signal, and
the terms $\{C^{\rm
    HI}_{\ell}[\xb b](\Delta \nu)\}^2/N_E$  and $\{\Delta
    C_{\ell}(\Delta \nu)\}^2_{sys}$ are respectively  the sample
variance and system noise contributions to the total variance. 

The fact that for $\ell=1476$ the measured $C_{\ell}(\Delta \nu)$ is
consistent with noise, and the signal is not detected allows us to use
eq. (\ref{eq:det}) to place an upper limit on $\xb b$.  The
filter $\tilde{F}(\tau)$ that has been used to remove the oscillatory
pattern in the residual also affects the signal. We have applied the
same filter to $C^{\rm HI}_{\ell}[\xb b](\Delta \nu)$ 
(Figure \ref{fig:xh1b}) and used this in eq. (\ref{eq:det}). The filtered
signal is maximum at $\Delta \nu=0$ and we use this data point to
place a $3\sigma$ upper limit on $\xb b$. 
A value of $\xb b$ greater than $7.95$ would have
been detected in our observation, and is therefore  ruled out 
at the $3\sigma$ level.
Our upper limit  is
around $330$ times larger than the value that we have estimated based
on results from quasar absorption spectra which imply $\xb=2.45 \times
10^{-2}$ and the assumption that $b=1$.  The HI signal should, in
principle, be detectable in  observations that are a few
hundred times more sensitive than the one that has been analyzed
here. 

\section{Acknowledgment}
AG would like to thank Prakash Sarkar, Suman Majumder, Prasun Dutta,
Tapomoy Guha Sarkar for many helpful discussions. AG would like to thank
Subhashis Roy of NCRA, Pune for useful discussions during the data
reduction stage of the analysis. AG would like to acknowledge Council
of Scientific and Industrial Research (CSIR), India for providing
financial support. SSA would like to acknowledge C.T.S,
I.I.T. Kharagpur for the use of its facilities. The data used in this
paper were obtained using GMRT. The GMRT is run by the National Centre
for Radio Astrophysics of the Tata Institute of Fundamental
Research. We thank the GMRT staff for making these observations
possible.

\end{document}